\documentclass[aps,prl,floatfix,superscriptaddress,preprintnumbers,nofootinbib]{revtex4-1}
\usepackage[utf8]{inputenc}
\usepackage{epsfig,latexsym,cancel,amssymb,amsmath,verbatim,mathrsfs}
\usepackage{color}
\usepackage{subfigure}
\usepackage{graphicx}
\newcommand{\be}{\begin{equation}}
\newcommand{\ee}{\end{equation}}
\newcommand{\bea}{\begin{eqnarray}}
\newcommand{\eea}{\end{eqnarray}}

\usepackage{tikz}
\usetikzlibrary{arrows,shapes}
\usetikzlibrary{trees}
\usetikzlibrary{matrix,arrows} 				
\usetikzlibrary{positioning}				
\usetikzlibrary{calc,through}				
\usetikzlibrary{decorations.pathreplacing}  
\usepackage{pgffor}							

\usetikzlibrary{decorations.pathmorphing}	
\usetikzlibrary{decorations.markings}
\tikzset{
    vector/.style={decorate, decoration={snake}, draw},
	provector/.style={decorate, decoration={snake,amplitude=2.5pt}, draw},
	antivector/.style={decorate, decoration={snake,amplitude=-2.5pt}, draw},
    fermion/.style={draw=black, postaction={decorate},
        decoration={markings,mark=at position .55 with {\arrow[draw=black]{>}}}},
    fermionbar/.style={draw=black, postaction={decorate},
        decoration={markings,mark=at position .55 with {\arrow[draw=black]{<}}}},
    fermionnoarrow/.style={draw=black},
    gluon/.style={decorate, draw=black,
        decoration={coil,amplitude=4pt, segment length=5pt}},
    scalar/.style={dashed,draw=black, postaction={decorate},
        decoration={markings,mark=at position .55 with {\arrow[draw=black]{>}}}},
    scalarbar/.style={dashed,draw=black, postaction={decorate},
        decoration={markings,mark=at position .55 with {\arrow[draw=black]{<}}}},
    scalarnoarrow/.style={dashed,draw=black},
    electron/.style={draw=black, postaction={decorate},
        decoration={markings,mark=at position .55 with {\arrow[draw=black]{>}}}},
	bigvector/.style={decorate, decoration={snake,amplitude=4pt}, draw},
}

\tikzstyle{block} = [draw, rectangle, 
    minimum height=3em, minimum width=6em]
\begin{document}

\title{Common origin of radiative neutrino mass, dark matter and leptogenesis in scotogenic Georgi-Machacek model}

\author{Shao-Long Chen}
\email[E-mail: ]{chensl@mail.ccnu.edu.cn}
\affiliation{Key Laboratory of Quark and Lepton Physics (MoE) and Institute of Particle Physics, Central China 
Normal University, Wuhan 430079, China}
\affiliation{Center for High Energy Physics, Peking University, Beijing 100871, China}

\author{Amit Dutta Banik}
\email[E-mail: ]{amitdbanik@mail.ccnu.edu.cn}
\affiliation{Key Laboratory of Quark and Lepton Physics (MoE) and Institute of Particle Physics, Central China 
Normal University, Wuhan 430079, China}

\author{Ze-Kun Liu}
\email[E-mail: ]{zekunliu@mails.ccnu.edu.cn}
\affiliation{Key Laboratory of Quark and Lepton Physics (MoE) and Institute of Particle Physics, Central China 
Normal University, Wuhan 430079, China}

\date{\today}

\begin{abstract}
We explore the phenomenology of the Georgi-Machacek model extended with two Higgs doublets and vector fermion doublets invariant under $SU(2)_L \times U(1)_Y\times \mathcal {Z}_4 \times \mathcal {Z}_2$. The $\mathcal {Z}_4$ symmetry is broken spontaneously while the imposed $\mathcal {Z}_2$ symmetry forbids triplet fields to generate any vacuum expectation value and leading to an inert dark sector providing a viable candidate for dark matter and generate neutrino mass radiatively.  Another interesting feature of the model is leptogenesis arising from decay of vector-like fermions. A detailed study of the model is pursued in search for available parameter space consistent with the theoretical and experimental observations for dark matter, neutrino physics, flavor physics, matter-antimatter asymmetry in the Universe.   
\end{abstract}

\maketitle

\subsection{Introduction}
Evidences from cosmology and astrophysics claim that about a quarter of the Universe is made up of dark matter~\cite{Aghanim:2018eyx}. However, the nature of dark matter remains unknown as the Standard Model (SM) of particle physics fails to provide a viable dark matter (DM) candidate. Despite being celebrated as the most successful theory after the discovery of Higgs boson at collider experiments, various theories beyond the SM are proposed in order to explain the dark matter. Simple extensions of SM with DM candidate are probed considering the stability of dark matter is protected by an additional discrete symmetry (such as $\mathcal {Z}_2$ or $\mathcal {Z}_3$ etc.). Apart from the dark matter, the origin of neutrino mass also remains unexplained by the SM of particle physics despite various neutrinos oscillation experiments has confirmed that neutrinos are massive~\cite{PhysRevD.98.030001}. This discrepancy in SM also calls for theories beyond the SM. Neutrino masses can indeed be generated by various see-saw mechanisms~\cite{Minkowski:1977sc,GellMann:1980vs,Mohapatra:1979ia,Schechter:1980gr,Mohapatra:1980yp,Wetterich:1981bx,Schechter:1981cv, Hambye:2003ka,Antusch:2007km,Foot:1988aq,Chen:2009vx} at tree level  with new heavy fermions and scalar fields, which can also generate matter-antimatter asymmetry in the Universe through the leptogenesis mechanism~\cite{Fukugita:1986hr,Buchmuller:2004nz,Buchmuller:2005eh,Davidson:2008bu,An:2009vq}. There are alternative ways to generate neutrino masses and matter-antimatter asymmetry where neutrino mass is generated radiatively via loop involving right-handed neutrinos~\cite{Ma:2006km,Kashiwase:2012xd,Kashiwase:2013uy,Hugle:2018qbw} or new fields~\cite{Lu:2016dbc,Cao:2017xgk,Zhou:2017lrt,Gu:2018kmv,DuttaBanik:2020vfr,Lineros:2020eit}, which can also provide viable  dark matter candidates. 

Among various extensions beyond the SM, two Higgs doublet model (2HDM) is one of the most simplest extensions where an additional scalar doublet similar to SM Higgs doublet is added~\cite{HABER1979493,HALL1981397,PhysRevD.41.3421,Branco:2011iw}. In conventional models of 2HDM, the newly added doublet is assumed to be odd under a $\mathcal {Z}_2$ symmetry which is broken spontaneously after electroweak symmetry is broken. After spontaneous symmetry breaking (SSB) doublet scalar fields obtain vacuum expectation values (VEVs) and mix up, resulting new physical Higgs particles. In the present work, we explore the phenomenology of a 2HDM extended Georgi-Machacek (GM) model~\cite{Georgi:1985nv,Chanowitz:1985ug,Gunion:1989ci}. In the GM model, new scalar triplets with hypercharge $Y=0$ and $Y=1$ are added to the SM scalar sector in a way that the electroweak $\rho$-parameter remains unaffected by simply assuming both the triplets develop same VEVs after spontaneous breaking of symmetry~\cite{Hartling:2014zca,Chiang:2015kka,Chiang:2015rva,Chiang:2018cgb,Das:2018vkv}. The hypercharge $Y=1$ triplet in GM model also produces tiny neutrino masses at tree level. We propose an extension of the GM model with an additional scalar doublet and new vector-like fermions charged under $SU(2)_L \times U(1)_Y\times \mathcal{Z}_4 \times \mathcal{Z}_2$ symmetry. After spontaneous breaking of the symmetry, both the doublet scalar fields develop VEVs resembling the usual 2HDM (which breaks $\mathcal{Z}_4$ symmetry of the model) while the triplet fields and vector fermion doublets remain protected by the imposed $\mathcal {Z}_2$ which constitutes the dark sector. The neutrino mass generation at tree level via triplet field is prohibited. However, new vector-like fermions provide a window to radiatively generate neutrino mass in one-loop level. 
In addition, the lightest neutral scalar particle arising from 
mixing between triplet fields being charged under the remnant $\mathcal {Z}_2$, is stable and serves as a dark matter candidate. Finally, CP violating decays of heavy vector fermions can explain the matter-antimatter asymmetry in the Universe via leptogenesis. With new scalar and fermion fields the modified GM model is referred as scotogenic Georgi-Machacek (sGM) model.

The paper is organized as follows, we firstly describe the sGM model in detail. The phenomenologies of the model are then performed with considering various theoretical and experimental limits. Study of leptogenesis within the model is presented in the next section. Finally we conclude the paper.   
             
\subsection*{sGM Model}
The scalar sector of the GM model~\cite{Georgi:1985nv} contains one Higgs doublet $\phi$ with $Y=1/2$, one complex triplet $\Delta$ with $Y=1$ and a real scalar triplet $T$ with $Y=0$ fields. 
\be
\begin{aligned}
\phi = \left(
\begin{array}{c}
\phi^+ \\
\phi^0
\end{array}\right),~
\Delta = \left(
\begin{array}{cc}
\frac{D^{+}}{\sqrt{2}} & -D^{++}\\
D^{0} & -\frac{D^{+}}{\sqrt{2}}
\end{array}\right),~
T = \left(
\begin{array}{cc}
\frac{T^0}{\sqrt{2}} & -T^+\\
-T^- & -\frac{T^0}{\sqrt{2}}
\end{array}\right), \label{par}
\end{aligned}
\ee
where the neutral components are parametrized as  
\be
\begin{aligned}
\phi^0 =\frac{1}{\sqrt{2}}(\phi_r+v_\phi +i\phi_i),\quad 
D^{0} =\frac{1}{\sqrt{2}}(D_r^0+i D_i^0)+v_\Delta,\quad 
T^0 = T_r+v_T,   \label{neut}
\end{aligned}
\ee
with $v_\phi$, $v_\Delta$ and $v_T$ being the VEVs for $\phi^0$, $D^{0}$ and $T^0$, respectively. The most general form of the Higgs potential, invariant under the $SU(2)_L\times U(1)_Y$ gauge symmetry, is given by
\bea
V(\phi,\Delta,T)&=& m_\phi^2(\phi^\dagger \phi)+m_\Delta^2\text{Tr}(\Delta^\dagger\Delta)+
\frac{m_T^2}{2}\text{Tr}(T^2)\nonumber\\
&&+\mu_1\phi^\dagger T\phi +\mu_2 [\phi^T(i\tau_2) \Delta^\dagger \phi+\text{h.c.}] +
\mu_3\text{Tr}(\Delta^\dagger \Delta T)
+\lambda_\phi (\phi^\dagger \phi)^2 \nonumber\\
&&
+\rho_1[\text{Tr}(\Delta^\dagger\Delta)]^2+\rho_2\text{Tr}(\Delta^\dagger \Delta\Delta^\dagger \Delta)
+\rho_3\text{Tr}(T^4)
+\rho_4 \text{Tr}(\Delta^\dagger\Delta)\text{Tr}(T^2)
+\rho_5\text{Tr}(\Delta^\dagger T)\text{Tr}(T \Delta)\nonumber\\
&&+\kappa_1\text{Tr}(\Delta^\dagger \Delta)\phi^\dagger \phi+\kappa_2 \phi^\dagger \Delta\Delta^\dagger \phi
+\frac{\kappa_3}{2}\text{Tr}(T^2)\phi^\dagger \phi 
+\kappa (\phi^\dagger \Delta T \tilde{\phi} + \text{h.c.}), \label{pot_gen}
\label{pot1}
\eea
where $\tilde{\phi}=i\tau_2\phi^*$. 
In the above Eq.~(\ref{pot_gen}), $\mu_2$ and $\kappa$ can be complex. 

In the present work we extend the model in the framework of two Higgs doublets with 
vector-like fermions (VLF's),
\be 
\Sigma_L\equiv (\Sigma_L^0~~~~ \Sigma_L^-)^T, \hskip 5mm
\Sigma_R\equiv (\Sigma_R^0~~~~ \Sigma_R^-)^T\, .
\ee
We consider a $\mathcal {Z}_4$ symmetry associated with the model instead of standard $\mathcal {Z}_2$ symmetry in 2HDM. 
In addition, we introduce a $\mathcal {Z}_2$ symmetry under which both the 
triplets and vector fermion doublets are odd which constitute the dark sector.
Charges of different particles and fields are given in 
Table~\ref{tab:1}.
\begin{table}[htb]
 \begin{center}
 \begin{tabular}{|c| c| c| c|c|c|} 
 \hline
 \hline
Fields & $SU(3)_c$ & $SU(2)_L$ & $U(1)_Y$ & $~\mathcal{Z}_4~$ & $~\mathcal{Z}_2~$\\ [0.5ex] 
 \hline\hline
 $T$ & 1 & 3 & 0 & 1 & - \\ 
 \hline
 $\Delta$ & 1 & 3 & 1 & 1 & - \\
 \hline
 $\phi_1$ & 1 & 2 & $ 1/2 $ & -i & + \\
 \hline
 $\phi_2$ & 1 & 2 & $ 1/2 $ & i & + \\
 \hline
 $\Sigma_{L,R}$& 1 & 2 & $ -1/2 $ & 1 & -\\
 \hline
 $Q_L$& 3 & 1 &  1/6 & 1 & + \\
 \hline
 $u_R$& 3 & 1 &  2/3 & i & + \\
 \hline
 $d_R$& 3 & 1 & -1/3 & -i & + \\
  \hline
  $L_L$& 1 & 2 &  -1/2 & 1 & + \\
 \hline
 $e_R$& 3 & 1 & -1 & -i & + \\
 \hline
 \hline
 \end{tabular}
\end{center}
\caption{Charge assignments of the fields in the sGM model.}
\label{tab:1}
\end{table} 

The general form of the Higgs potential, invariant under the $SU(2)_L\times U(1)_Y \times \mathcal {Z}_4 \times \mathcal {Z}_2$ symmetry, is 
parametrized as
\bea
V({\rm 2HDM},\Delta,T)&=&V_{\rm 2HDM}+m_\Delta^2\text{Tr}(\Delta^\dagger\Delta)+\frac{m_T^2}{2}\text{Tr}
(T^2) \nonumber\\
&&
+\rho_1[\text{Tr}(\Delta^\dagger\Delta)]^2+\rho_2\text{Tr}(\Delta^\dagger \Delta\Delta^\dagger \Delta)
+\rho_3\text{Tr}(T^4)
+\rho_4 \text{Tr}(\Delta^\dagger\Delta)\text{Tr}(T^2)
+\rho_5\text{Tr}(\Delta^\dagger T)\text{Tr}(T \Delta)\nonumber\\
&&+\sum\limits_{a=1,2} \left( \kappa_{1a}\text{Tr}(\Delta^\dagger \Delta)\phi_a^\dagger \phi_a+
\kappa_{2a} \phi_a^\dagger \Delta\Delta^\dagger 
\phi_a
+\frac{\kappa_{3a}}{2}\text{Tr}(T^2)\phi_a^\dagger \phi_a \right) \nonumber\\
&&+\kappa_{41} (\phi_1^\dagger \Delta T \tilde{\phi}_2 + \text{h.c.}) + \kappa_{42} (\phi_2^\dagger \Delta T \tilde{\phi}_1 
+ \text{h.c.})  \,\, , \label{pot_gen3}
\eea

where $\tilde{\phi}_{1,2}\equiv i\tau_{2}\phi_{1,2}^{*}$, and the 2HDM potential is given by 
\bea
V_{\rm 2HDM} &=&
m^2_{11}\, \phi_1^\dagger \phi_1
+ m^2_{22}\, \phi_2^\dagger \phi_2 
+ \frac{\lambda_1}{2} \left( \phi_1^\dagger \phi_1 \right)^2
+ \frac{\lambda_2}{2} \left( \phi_2^\dagger \phi_2 \right)^2
+ \lambda_3\, \phi_1^\dagger \phi_1\, \phi_2^\dagger \phi_2 \nonumber \\  &&
+ \lambda_4\, \phi_1^\dagger \phi_2\, \phi_2^\dagger \phi_1
+ \frac{\lambda_5}{2} \left[
\left( \phi_1^\dagger\phi_2 \right)^2
+ \left( \phi_2^\dagger\phi_1 \right)^2 \right].
\label{treepot}
\eea

One interesting aspect of the choice $\mathcal{Z}_4$ is that it forbids interaction terms $\phi_1^\dagger \Delta T \tilde{\phi}_1$ and $\phi_2^\dagger \Delta T \tilde{\phi}_2$ in Eq.~(\ref{pot_gen3}) and the soft breaking term $m_{12}^2\phi_1^\dagger\phi_2$ in 2HDM potential (Eq.~(\ref{treepot})). However new interactions in Eq.~(\ref{pot_gen3}) are allowed with coupling $\kappa_{41,42}$ which play significant role in generation of neutrino mass, as we will show later. The $\mathcal{Z}_4$ symmetry is broken spontaneously as the doublet fields $\phi_1$,  $\phi_2$ acquires VEVs. However, triplet fields in the potential expression of Eq.~(\ref{pot_gen3}) preserves the $\mathcal {Z}_2$ symmetry ($T\rightarrow -T$ and $\Delta\rightarrow -\Delta$ following Table~\ref{tab:1}) and does not acquire any VEV. Therefore, this remnant $\mathcal {Z}_2$ symmetry provides feasible candidates for dark matter arising from the mixing between neutral components of triplet fields in the model. After spontaneous breaking of symmetry we get new physical scalar in visible 2HDM sector and a dark sector originating from triplet scalars. The neutral scalar particles and singly charged particles of the triplets $T$ and $\Delta$ mix with each other and provides two neutral physical scalars and two physical charged scalars in dark sector. However, since the VEV of $\Delta$ field is zero due to the residual symmetry $\mathcal {Z}_2$, the neutrino mass is vanishing at tree level in the model. This issue can be resolved by the newly added vector-like fermions which also respects the unbroken $\mathcal {Z}_2$ symmetry.

Gauge invariant Yukawa interactions of the fermions with triplet scalars in the present framework are given as
\be
\mathcal{L}= M_{\Sigma}\bar{\Sigma}_L\Sigma_R + y \bar{L}_L^c i \tau_2 \Delta 
\Sigma_L + \lambda \bar{L}_LT\Sigma_R+ \text{h.c.} \,\, .
\label{int}
\ee
 We will later show that these new interaction terms provide necessary ingredients to radiatively generate neutrino masses in one-loop.

\subsubsection*{Scalar Sector}
After spontaneous symmetry breaking, Higgs fields $\phi_1$ and $\phi_2$ acquire vacuum expectation values $v_1$ and $v_2$, such that $v=\sqrt{v_1^2+v_2^2}=246$ GeV. The visible sector is identical to 2HDM which contains two neutral physical scalars $(h, H)$, one pseudo-scalar particle $A$ and a pair of charged scalar $H^\pm$. Conditions for the minimization of the potential are
\bea
m_{11}^2= \frac{ -\lambda_1 v_1^2 - (\lambda_3+\lambda_4+\lambda_5)v_2^2}{2}\,\, , \nonumber \\
m_{22}^2= \frac{ -\lambda_2 v_2^2 - (\lambda_3 +\lambda_4+\lambda_5)v_1^2}{2}\, .
\label{conditions}
\eea

Different couplings $\lambda_i (i=1,...,5)$ are expressed in terms of physical masses $m_h,~m_H,~m_A,~m_{H^\pm}$ and parameters $\alpha,~\beta $,
\bea
\label{e:l1}
\lambda_1 &=& \frac{1}{v^2 c^2_\beta}~\Big(c^2_\alpha m^2_H + s^2_\alpha m^2_h   \Big),\\
\label{e:l2}
\lambda_2 &=& \frac{1}{v^2 s^2_\beta}~\Big(s^2_\alpha m^2_H + c^2_\alpha m^2_h \Big),\\
\label{e:l4}
\lambda_4 &=& \frac{1}{v^2}~(m^2_A - 2 m^2_{H^+}),\\
\label{e:l5}
\lambda_5 &=& - \frac{m^2_A}{v^2}, \\
\label{e:l3}
\lambda_3 &=& \frac{1}{v^2 s_\beta c_\beta}\left((m^2_H - m^2_h)s_\alpha c_\alpha + m^2_A s_\beta c_\beta \right) - 
\lambda_4\, ,
\label{coup}
\eea
where we denote $s_{\alpha,\beta}=\sin{\alpha},\sin{\beta},~c_{\alpha,\beta}=\cos{\alpha},\cos{\beta}$, with $\tan{\beta}=v_{2}/v_{1}$. Here $m_h=125$ GeV is the mass of SM Higgs and $m_H$ is denoted as the mass of heavy Higgs boson. After the SSB, the scalar fields $T$ and $\Delta$ acquire zero VEVs which resemble an unbroken residual $\mathcal {Z}_2$ symmetry. Therefore, the scalar fields $T$ and $\Delta$ are inert in nature. Mass terms for different inert scalar particles are given as follows
\bea
m_{T^0}^2 &=& M_T^2 + \frac{1}{2}\kappa_{31} v_1^2 +\frac{1}{2}\kappa_{32} v_2^2 = m_{T^+}^2 \, ,\nonumber \\
m_{D^0}^2 &=& M_{\Delta}^2 + \frac{1}{2}(\kappa_{11} + \kappa_{21})v_1^2 + \frac{1}{2}(\kappa_{12} + 
\kappa_{22})v_2^2 = m_{A^0}^2\, , \nonumber \\
m_{D^{++}}^2 &=& M_{\Delta}^2 + \frac{1}{2}\kappa_{11} v_1^2 + \frac{1}{2}\kappa_{12} v_2^2 \, ,\nonumber \\
m_{D^{+}}^2 &=& M_{\Delta}^2 + \frac{1}{2}(\kappa_{11} + \frac{1}{2}\kappa_{21})v_1^2 + \frac{1}{2}(\kappa_{12} 
+ \frac{1}{2}\kappa_{22})v_2^2 \, , \nonumber \\
m_{TD}^{2}&=& -\left(\frac{\kappa_{41} v_1 v_2}{2} + \frac{\kappa_{42} v_1 v_2}{2}\right)\, , \nonumber \\ 
m_{T^+D^+}^{2}&=& \frac{\kappa_{41} v_1 v_2}{2\sqrt{2}} + \frac{\kappa_{42} v_1 v_2}{2\sqrt{2}}\, .
\label{massterm}   
\eea  
The mass of inert pseudo-scalar is denoted as $m_{A^0}$ and $m_{D^{++}}$ is the mass of doubly charged scalar. Neutral parts of both the triplets mix with each other resulting two new physical neutral scalars $S_{1,2}$ and the mass matrix is given as 
\be
\begin{aligned}
M_{neutral}^2= \left(
\begin{array}{cc}
m_{T^0}^2 & m_{TD}^2\\
m_{TD}^2 & m_{D^0}^2
\end{array}\right).~
\end{aligned}
\ee
We define a mixing angle $\gamma$ between these two inert scalar fields such that
\bea 
S_1=T^0~\cos \gamma - D_r^0~\sin \gamma \, , \nonumber \\
S_2=T^0~\sin \gamma + D_r^0~\cos \gamma \, . 
\label{neutral}
\eea   
Masses of new physical scalars are expressed as
\be
m^2_{S_1/S_2}=\frac{m_{T^0}^2+m_{D^0}^2}{2}\mp\frac{m_{D^0}^2-m_{T^0}^2}{2}\sqrt{1+x^2},
\label{mass1} 
\ee
where $x=\tan 2\gamma=\frac{2m_{TD}^2}{(m_{D^0}^2-m_{T^0}^2)}$.  Similarly, the charged parts also mix with each other and provides two physical charged scalars $S_{1,2}^+$. The mass matrix for the charged scalars is
\be
\begin{aligned}
M_{charged}^2= \left(
\begin{array}{cc}
m_{T^+}^2 & m_{T^+D^+}^2\\
m_{T^+D^+}^2 & m_{D^+}^2
\end{array}\right).~
\end{aligned}
\ee
Defining a new mixing angle $\delta$, we write physical charged scalars as
\bea 
S_1^+=T^+~\cos \delta - D^{+}~\sin \delta \, , \nonumber \\
S_2^+=T^+~\sin \delta + D^{+}~\cos \delta \, . 
\label{charged}
\eea   
Masses of physical charged scalars are 
\be
m^2_{S_1^+/S_2^+}=\frac{m_{T^+}^2+m_{D^+}^2}{2}\mp\frac{m_{D^+}^2-m_{T^+}^2}{2}\sqrt{1+y^2},
\label{mass2} 
\ee  
with $y=\tan 2\delta=\frac{2m_{T^+D^+}^2}{(m_{D^+}^2-m_{T^+}^2)}$.

\subsubsection*{Vacuum stability}
We adopt the criteria of copositivity of symmetric matrices to get the conditions of vacuum stability~\cite{Chakrabortty:2013mha}. Only the quartic terms in the potential should be considered, since these terms dominate at large field value. It is to be noted that vacuum stability conditions don't give any constraints to couplings $\kappa_{41}$ and $\kappa_{42}$. This is because we can make these couplings positive by applying a phase rotation of a field or field redefinition. We parameterize the fields as,
\begin{equation}
\begin{split}
&\phi_1^{\dagger} \phi_1=|h_1|^2 \qquad \phi_2^{\dagger} \phi_2=|h_2|^2 \qquad \Delta^{\dagger} \Delta=
|\delta|^2 \qquad T^{\dagger} T=|t|^2 \qquad \phi_1^{\dagger} \phi_2= f |h_1||h_2|e^{i\theta_1}\\
&\Delta^{\dagger} T =g |\delta||t|e^{i\theta_2} \qquad \Delta^{\dagger} \phi_1=m|\delta||\phi_1|e^{i\theta_3} 
\qquad \Delta^{\dagger} \phi_2=n|\delta||\phi_2|e^{i\theta_4}
\end{split}
\end{equation}
where $f, g, m, n \in [0,1]$.  According to the definition above, considering the quartic terms out and ignoring 
$\kappa_{41}$ and $\kappa_{42}$ terms, the potential of Eq.~(\ref{pot_gen3}) can be expressed as
\begin{equation}
\begin{split}
V(h_1,h_2,\delta,t) =&\frac{\lambda_1}{2} h_1^4 + \frac{\lambda_2}{2} h_2^4 + \lambda_3\, h_1^2 h_2^2 + 
\lambda_4\,\rho^2 h_1^2 h_2^2 + \frac{\lambda_5}{2} f^2\, \cos 2\theta_1\,  h_1^2 h_2^2 + (\rho_1 + \rho_2)\, \delta^4 \\
& + \rho_3\, t^4 + (\rho_4+\rho_5\, g^2)\delta^2 t^2 + (\kappa_{11} + \kappa_{21} m^2)\delta^2 h_1^2  
+ \frac{\kappa_{31}}{2} t^2 h_1^2 \\
&+ (\kappa_{11} + \kappa_{22} n^2)\delta^2 h_2^2  + \frac{\kappa_{32}}{2} t^2 h_2^2
\end{split}
\label{vac2}
\end{equation}
The symmetric matrix of quartic couplings can be represented in basis $(h_1^2, h_2^2, \delta^2, t^2)$ as
\begin{equation}
M=
\begin{pmatrix}
\frac{1}{2}\lambda_1 & \frac{1}{2} \lambda_{345} &\frac{1}{2}\left(\kappa_{11} + m^2 \kappa_{21} \right) &\frac{1}{4}\kappa_{31} \\
 & \frac{1}{2}\lambda_2 &\frac{1}{2}\left(\kappa_{12} + n^2 \kappa_{22} \right) &\frac{1}{4} \kappa_{32}\\
 &  &\rho_1 + \rho_2 &\frac{1}{2} \left(\rho_4 + g^2 \rho_5 \right)\\
 &  & &\rho_3 \\
\end{pmatrix}
\label{matrix}
\end{equation}
where\ $\lambda_{345}=\lambda_3 + f^2 (\lambda_4 - |\lambda_5|)$. If $(\lambda_4 - |\lambda_5|) \ge 0$, the minimum of the potential is obtained for  $f=0$, whereas for $(\lambda_4 - |\lambda_5|) < 0$ the minimum obtained assuming $f=1$. Similar convention can be employed to parameters $g, m, n$  to apply the copositivity criteria upon the matrix $M$. Copositive conditions for which vacuum of the scalar potential in Eq.~(\ref{vac2}) 
becomes stable are:
\begin{equation}
\begin{split}
&\lambda_1 \ge 0 , \quad \lambda_2 \ge 0, \quad \rho_1+\rho_2 \ge 0, \quad \rho_3 \ge 
0, \\
&\kappa_{11}+\kappa_{21} +\sqrt{2\lambda_1(\rho_1 + \rho_2)} \ge 0, \\
&\kappa_{11}+\sqrt{2\lambda_1(\rho_1 + \rho_2)} \ge 0, \\
&\kappa_{12}+\kappa_{22} +\sqrt{2\lambda_1(\rho_1 + \rho_2)} \ge 0, \\
&\kappa_{12}+\sqrt{2\lambda_1(\rho_1 + \rho_2)} \ge 0, \\
&\lambda_3 + \lambda_4 - |\lambda_5| + \sqrt{\lambda_1 \lambda_2} \ge 0, \\
&\lambda_3 + \sqrt{\lambda_1 \lambda_2} \ge 0, \\
&8\lambda_1\rho_3 - \kappa_{31}^2 \ge 0, \\
&8\lambda_2\rho_3 - \kappa_{32}^2 \ge 0, \\
&8(\lambda_3+\lambda_4- |\lambda_5|)\rho_3 - \kappa_{31} \kappa_{32}+ 2\sqrt{(8\lambda_1\rho_3-
\kappa_{31}^2)(8\lambda_2\rho_3-\kappa_{32}^2)} \ge 0 ,\\
&8\lambda_3\rho_3 - \kappa_{31} \kappa_{32}+ 2\sqrt{(8\lambda_1\rho_3-\kappa_{31}^2)(8\lambda_2\rho_3-
\kappa_{32}^2)} \ge 0, 
\end{split}
\label{vsc}
\end{equation}
where we used a theorem of copositivity criteria to get copositive matrix conditions~\cite{PING1993109}. The conditions mentioned above are derived assuming $\rho_4 \ge 0$ and $\rho_5 \ge 0$.

In addition, scalar and fermion couplings must also remain within the perturbative limit for which following conditions must be satisfied
\bea
\lambda_i,\kappa_{1i,2i,3i,4i},\rho_i<4\pi,\hskip 5 mm \lambda,y<\sqrt{4\pi}\, .
\eea

\subsubsection*{Neutrino mass}

\begin{figure}[!ht]
\centering
\begin{tikzpicture}[line width=1.5 pt, scale=1.3, >=latex]
	\draw[fermion] (-2,0)--(-1,0);
	\draw[fermion] (-1,0)--(0,0);
    \draw[scalar] (-1,0) arc (180:90:1);
    \draw[scalarbar] (1,0) arc (0:90:1);
\begin{scope}[shift={(90:1)}]
    \draw[scalarbar] (0,0)--(45:1.2);
    \draw[scalarbar] (0,0)--(135:1.2);
    \node at (35:1.2) {$\phi_2$};
    \node at (145:1.2) {$\phi_1$};
\end{scope}
	\draw[fermion] (0,0)--(1,0);
	\draw[fermionbar] (1,0)--(2,0);
	\node at (0,0) {$\times$};
	\node at (-2,-0.3) {$\nu$};
	\node at (-0.5,-0.3) {$\Sigma_R$};
	\node at (0.5,-0.3) {$\Sigma_L$};
	\node at (2,-0.3) {$\nu$};
	\node at (45:1.3) {$\Delta$};
	\node at (135:1.3) {$T$};
\end{tikzpicture}
\caption{One-loop neutrino mass in sGM model.}
\label{diag}
\end{figure}
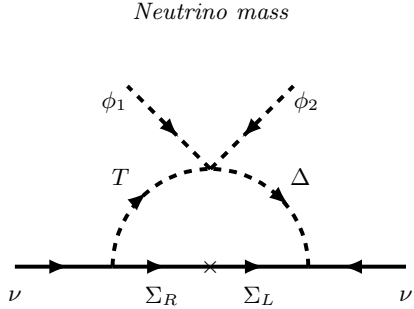
The neutrino masses are generated via one-loop diagram as shown in Fig.~\ref{diag}. The Yukawa interaction terms in Eq.~(\ref{int}) and scalar terms with $\kappa_{41}$ and $\kappa_{42}$ appearing in Eq.~(\ref{pot_gen3}) are responsible for generation of light neutrino mass. The $\mathcal {Z}_4$ symmetry in the model assures that two different scalar doublets are necessary in order to generate tiny neutrino mass in one-loop. 

\begin{figure}[!ht]
\centering
\begin{tikzpicture}[line width=1.5 pt, scale=1.3, >=latex]
	\draw[fermion] (-1,0)--(0,0);
	\draw[fermion] (0,0)--(1,0);
	\draw[scalarnoarrow] (0,0) arc (180:0:1);
	\draw[fermion] (1,0)--(2,0);
	\draw[fermionbar] (2,0)--(3,0);
	\node at (1,0) {$\times$};
	\node at (-1,-0.3) {$\nu$};
	\node at (1,-0.3) {$\Sigma_k$};
	\node at (3,-0.3) {$\nu$};
	\node at (1,1.3) {$S_{1,2}/S_{1,2}^+$};
\end{tikzpicture}
\caption{Neutrino mass from self energy diagrams.}
\label{fig2}
\end{figure}
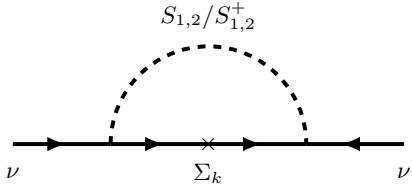
Neutrino mass in the present model can further be realized as self-energy corrections. However, there will be contributions from two different diagrams involving neutral scalars and charged scalars as shown in Fig.~\ref{fig2}. The expression of neutrino mass is given as \begin{eqnarray}\nonumber
({\cal M}_\nu)_{ij} &=& \cos \gamma \sin \gamma \sum_{k=1}^3 {[y_{ik} \lambda_{jk} + \lambda_{ik}y_{jk}]\over 
32 \pi^2}M_{\Sigma_k}\left[ {m_{S_1}^2 \over m_{S_1}^2 - M_{\Sigma_k}^2} \ln {m_{S_1}^2 \over M_{\Sigma_k}^2} -
{m_{S_2}^2 \over m_{S_2}^2 - M_{\Sigma_k}^2} \ln {m_{S_2}^2 \over M_{\Sigma_k}^2} \right] \\&&
- \cos \delta \sin \delta \sum_{k=1}^3 {[y_{ik} \lambda_{jk} + \lambda_{ik}y_{jk}]\over 32 \pi^2}M_{\Sigma_k}
\left[ {m_{S_1^+}^2 \over m_{S_1^+}^2 - M_{\Sigma_k}^2} \ln {m_{S_1^+}^2 \over M_{\Sigma_k}^2} -
{m_{S_2^+}^2 \over m_{S_2^+}^2 - M_{\Sigma_k}^2} \ln {m_{S_2^+}^2 \over M_{\Sigma_k}^2} \right]
\ .
\label{nmass}
\end{eqnarray}

It is to be noted that, in the present scenario if the mixing angle between neutral physical scalar $\sin\gamma \rightarrow 0$,  then mixing angle between charged scalars ($\sin\delta$) also becomes zero. The conditions for which those mixing angles become zero are
\be
\kappa_{41}=\kappa_{42}=0\, ,\hskip 5mm \kappa_{41}=-\kappa_{42}\, . 
\ee
Therefore, in order to generate tiny neutrino mass one must have $\kappa_{41,42}\neq 0$ and $\kappa_{41} \neq -\kappa_{42}$.

We  consider vector fermions to be much heavier than scalar particles $M_{\Sigma}\gg m_{S_{1,2}},m_{S_{1,2}^+}$, and further assume ${m_{S_1}^2}\simeq{m_{S_2}^2}$, ${m_{S_1^+}^2}\simeq{m_{S_2^+}^2}$. With the above simplified choice, following Eqs.~(\ref{massterm}), (\ref{mass1}) and Eq.~(\ref{mass2}) we can rewrite the neutrino mass matrix elements as
\begin{eqnarray}
({\cal M}_\nu)_{ij} &=& \left(\frac{\kappa_{41}+\kappa_{42}}{2}\right)v_1v_2\left[ \sum_{k=1}^3 {[y_{ik} 
\lambda_{jk} + \lambda_{ik}y_{jk}]\over 32 \pi^2}
\frac{1}{M_{\Sigma_k}} \ln {m_{S_1}^2 \over M_{\Sigma_k}^2}
+ \sum_{k=1}^3 {[y_{ik} \lambda_{jk} + \lambda_{ik}y_{jk}]\over 32 \pi^2}
\frac{1}{\sqrt{2}M_{\Sigma_k}} \ln {m_{S_1^+}^2 \over M_{\Sigma_k}^2} \right]
\ .
\label{nmass3}
\end{eqnarray}

Therefore, from Eq.~(\ref{nmass3}), one can realize seesaw like radiative neutrino mass expressed as
\begin{eqnarray}
{\cal M}_\nu &=& \frac{(2+\sqrt{2})(\kappa_{41}+\kappa_{42})v_1v_2}{128 \pi^2} (y \zeta^{-1} \lambda^T + \lambda \zeta^{-1} y^T)\,,
\label{nmass4}
\end{eqnarray}
where we assumed that the mass matrix of vector fermion $\zeta={\rm diag}\{\zeta_{1},\zeta_{2},\zeta_{3}\}$ and the diagonal matrix elements are given as 
\be
\zeta_i\equiv M_{\Sigma_i}\left[\ln {m_{S_1}^2 \over M_{\Sigma_i}^2} \right]^{-1}= M_{\Sigma_i}\left[\ln {m_{S_1^+}^2 \over M_{\Sigma_i}^2} \right]^{-1}\, .
\label{term}
\ee

Therefore, if one considers $(\kappa_{41}+\kappa_{42})\sim \mathcal{O}(1)$, $y\sim\lambda\simeq \mathcal{O}(10^{-2})$, 
$\tan\beta\sim1$ and $M_{\Sigma}\sim \mathcal{O}(10^{10})$ GeV,  we get neutrino mass scale at $m_\nu \simeq \mathcal{O}(0.05)$ eV
for TeV scale triplets in sGM framework. 
The order of Yukawa coupling also controls parameters significant for the process of leptogenesis which will be 
discussed later. The mass matrix ${\mathcal M}_\nu$ can be diagonalized by 
Pontecorvo-Maki-Nakagawa-Sakata (PMNS) matrix $U$ 
 \be
{{\mathcal M}_{\nu}} =U^{*}\cdot\hat{\mathcal M}_{\nu}\cdot U^{\dagger}
\label{numass2}
\ee
where $\hat{\mathcal M}_{\nu}={\rm{diag}}(m_1,m_2,m_3)$.

\subsection*{Theory and Phenomenology of the sGM model}

\subsubsection*{Electroweak precision test observables}
We consider the generalised form of GM model (Eq.~(\ref{pot_gen3})) since it is well known that custodial symmetry in GM model is broken at the one-loop level by hypercharge interactions~\cite{Gunion:1990dt,Blasi:2017xmc,Keeshan:2018ypw}.
After symmetries of the scalar potential breaks of spontaneously, $\mathcal{Z}_2$ symmetry is partially conserved by triplet scalars 
and both $Y = 0$ and $Y = 1$ triplet fields remain inert.
At this stage (after SSB), Higgs doublets and new inert triplets doesn't have any impact on $\rho$ parameter 
and $\rho=1$ is satisfied at tree level. 
However, one-loop corrections parameter to $\rho$ parameter must be taken into account which can also be 
obtained in terms of   ${\bar{T}}=\alpha T=\Delta\rho$ due to new scalars and fermions. Firstly, let us consider the contribution from new fermions within the model. For a single generation of vector doublet fermions, there will be new contribution to ${\bar{T}}$ parameter which is given as~\cite{Cynolter:2008ea}
\bea
\begin{split}
{\bar{T}}^{\rm VLF}=-\frac{g_2^2}{8\pi^2 m_W^2}~\Pi (m_1,m_2)\, ,
\end{split}
\label{eq:tvlf}
\eea
where 
\bea
\begin{split}
\Pi(m_1,m_2)&=-\frac{1}{2} \left(m_1^2+m_2^2\right) \left(\text{div}+\log \left(\frac{\mu_{EW}^2}{m_1m_2}\right)\right)\\&
+m_1 m_2 \left(\text{div}+\frac{\left(m_1^2+m_2^2\right) \log \left(\frac{m_2^2}{m_1^2}\right)}{2 \left(m_1^2-m_2^2\right)}
+\log \left(\frac{\mu_{EW}^2}{m_1 m_2}\right)+1\right)\\&-\frac{1}{4} \left(m_1^2+m_2^2\right)
-\frac{\left(m_1^4+m_2^4\right) \log \left(\frac{m_2^2}{m_1^2}\right)}{4 
\left(m_1^2-m_2^2\right)}\, .
\end{split}
\label{eq:pi}
\eea
In the present work we include three vector fermion doublets. However, since vector fermions do not mix with each other  and mass of charged and neutral fermions are degenerate (i.e.; $M_{\Sigma^+}=M_{\Sigma^0}=M_{\Sigma}$ for a single generation), the contribution ${\bar{T}}^{\rm VLF}=0$.  

Let us now discuss how electroweak precision observables modify in presence of new scalars. In case of  2HDM within alignment limit $(\beta-\alpha)=\pi/2$, contribution to ${\bar{T}}$ parameter reads as 
~\cite{He:2001tp,Grimus:2007if,Grimus:2008nb}
\bea
{\bar{T}}^{\rm 2HDM} = \frac{g_2^2}{64\pi^2 m_W^2}\left(\xi\left(m_{H^\pm}^2,m_A^2\right)+
\xi\left(m_{H^\pm}^2,m_H^2\right)-\xi\left(m_A^2,m_H^2\right)\right),
\label{t2hdm}
\eea
where
\bea
\xi\left(x,y\right) = \begin{cases}                       
\frac{x+y}{2}-\frac{xy}{x-y}\ln\left(\frac{x}{y}\right), & \text{if $x\neq y$}.\\
0, & \text{if $x=y$}.\\
\end{cases}
\label{eq:fxy}
\eea
 From Eq.~(\ref{t2hdm}), it can be concluded that ${\bar{T}}^{\rm 2HDM}$ vanishes for $m_A=m_{H^\pm}$ or  $m_H=m_{H^\pm}$.  Therefore, if one considers above conditions, new contribution to ${\bar{T}}$ parameter  will arise from triplet scalar fields $T$ and $\Delta$ only. It is to noted that, in the present formalism there exists  mixing between neutral and charged components of the triplet fields, which must be taken into account to  calculate ${\bar{T}}$.  Considering the effects of mixing, additional contribution to ${\bar{T}}$ parameter is 
\bea
{\bar{T}^{new}} =\frac{g_2^2}{64\pi^2 m_{W}^2}\Big(s_{\delta}^2\xi(m_{D^{\pm\pm}}^2,m_{S_1^\pm}^2)+
c_{\delta}^2\xi (m_{D^{\pm\pm}}^2,m_{S_2^\pm}^2) + 
(c_{\delta}s_{\gamma}/\sqrt{2}+c_{\gamma} s_{\delta})^2 
\xi(m_{S_2^{\pm}}^2,m_{S_1}^2)\nonumber \\ +
(c_{\delta}c_{\gamma}/\sqrt{2}-s_{\gamma} s_{\delta})^2 
\xi (m_{S_2^{\pm}}^2,m_{S_2}^2)+
(s_{\delta}s_{\gamma}/\sqrt{2}-c_{\gamma} c_{\delta})^2 
\xi(m_{S_1^{\pm}}^2,m_{S_1}^2)  \nonumber \\ +
(s_{\delta}c_{\gamma}/\sqrt{2}+s_{\gamma} c_{\delta})^2 
\xi(m_{S_2}^2,m_{S_1^{\pm}}^2)
+s_{\delta}^2\xi(m_{S_1^\pm}^2,m_{A^0}^2)/2+
c_{\delta}^2\xi (m_{S_2^\pm}^2,m_{A^0}^2)/2\nonumber \\
-2s_{\delta}^2 c_{\delta}^2\xi (m_{S_2^\pm}^2,m_{S_1^\pm}^2)-
c_{\gamma}^2\xi (m_{S_2}^2,m_{A^0}^2)-s_{\gamma}^2\xi (m_{S_1}^2,m_{A^0}^2)\Big)\, .
\label{Tgm}
\eea
In the above Eq.~(\ref{Tgm}), $s_{\theta} =\sin\theta$ and $c_{\theta}=\cos\theta$ where $\theta=\gamma,~\delta$. Using the above expression, bounds on the mixing angles or mass splittings between scalar particles can be obtained in the present model. We use the value $T=\frac{\bar{T}^{new}}{\alpha} =0.07\pm0.12$~\cite{PhysRevD.98.030001} to constrain the model parameter space.

 \subsubsection*{Collider bounds}

In this section, we briefly discuss collider constraints on different visible and dark sector particles of the model. LEP excludes a new charged scalar of mass less than 80 GeV from charged scalar decay into $cs$ and $\nu \tau$ final state~\cite{Abbiendi:2013hk}. Similarly bound on charged fermion mass is 101.2 GeV~\cite{Abdallah:2003xe,Achard:2001qw} from the decay of charged fermion into $\nu W^\pm$ final state. Let us now consider the bounds obtained from LHC. In the present model we have singly charged scalars $S_{1,2}^\pm$ and one doubly charged scalar $D^{\pm\pm}$ in dark sector. These particles can contribute to the Higgs to diphoton decay process. The decay rate for the process $h\rightarrow \gamma \gamma$ is given as

\begin{eqnarray}
\label{eq:THM-h2gaga}
\Gamma(h \rightarrow\gamma\gamma)
& = & \frac{\alpha^2 G_F m_{h}^3}
{128\sqrt{2}\pi^3} \bigg| \sum_f N_c Q_f^2 g_{h f\bar{f}} 
A^h_{1/2}
(\tau_f) + g_{h W^+W^-} A^{h}_{1}(\tau_W) +  \frac{\lambda_{h H^\pm\,H^\mp}v}{2m^2_{H^\pm}} 
A^h_0(\tau_{H^{\pm}}) \nonumber \\
&& \hspace{1.5cm} + \frac{\lambda_{h S_1^\pm\,S_1^\mp}v}{2m^2_{S_1^\pm}} A^h_0(\tau_{S_1^{\pm}})
+ \frac{\lambda_{h S_2^\pm\,S_2^\mp}v}{2m^2_{S_2^\pm}} A^h_0(\tau_{S_2^{\pm}})
+ 4 {\frac{\lambda_{h D^{\pm\pm}D^{\mp\mp}}v}{2m^2_{D^{\pm\pm}}}
A^h_0(\tau_{D^{\pm\pm}})} \bigg|^2 \, .
\label{partial_width_htm}
\end{eqnarray}
Here $G_F$ is the Fermi coupling constant, $\alpha$ is the fine-structure constant, $N_c=3 (1)$ for quarks (leptons), $Q_f$ is the electric charge of the fermion in the loop, and $\tau_i=m_h^2/4m_i^2~(i=f,W,S_1^\pm,H^\pm,S_2^\pm,D^{\pm\pm})$. Couplings of SM Higgs $h$ with different charged scalars in Eq.~(\ref{partial_width_htm}) are listed in Appendix. The relevant loop functions are given by 
\begin{eqnarray}
A^h_{1/2}(\tau)&=& 2\left[\tau+(\tau-1)f(\tau)\right]\tau^{-2}, 
\label{eq:Afermion}\\
A^h_{1}(\tau)&=& -\left[2\tau^2+3\tau+3(2\tau-1)f(\tau)\right]\tau^{-2}, 
\label{eq:Avector}\\ 
A^h_{0}(\tau) &=& -[\tau -f(\tau)] \tau^{-2} \, ,
\label{eq:Ascalar}
\end{eqnarray}
 and the function $f(\tau)$ is given by
\begin{eqnarray}
f(\tau)=\left\{
\begin{array}{ll}  \displaystyle
\left[\sin^{-1}\left(\sqrt{\tau}\right)\right]^2, & (\tau\leq 1), \\
\displaystyle -\frac{1}{4}\left[ \log\left(\frac{1+\sqrt{1-\tau^{-1}}}
{1-\sqrt{1-\tau^{-1}}}\right)-i\pi \right]^2, \hspace{0.5cm} & (\tau>1) \, .
\end{array} \right. 
\label{eq:ftau} 
\end{eqnarray}
One can calculate the quantity called signal strength $R_{\gamma \gamma}$ given as
\bea
R_{\gamma \gamma} = \frac{\sigma(pp\rightarrow h)}
{\sigma(pp\rightarrow h)^{\rm SM}}
\frac{Br(h\rightarrow \gamma \gamma)}{Br(h\rightarrow \gamma \gamma)^{\rm SM}}
\label{diphoton}
\eea
Using the latest and future constraints from LHC on the $R_{\gamma \gamma}$ signal strength~\cite{Aaboud:2018xdt,Sirunyan:2018koj}, masses of charged particles in our model can be constrained. It is to be noted that for heavier masses of charged inert scalar particles with mass around TeV, one can safely work with 2HDM constraints. In the present framework, we consider only one Higgs doublet $\phi_2$ couples to the SM sector which resembles the standard type-I 2HDM. 
For type-I 2HDM, Higgs signal strength does not provide any limit on $\tan\beta$ with alignment limit $(\beta-\alpha)=\pi/2$. However, deviation from the alignment limit puts stringent bound on $\tan\beta$  for type-I 2HDM~\cite{Arcadi:2018pfo,Khachatryan:2014jba,Aad:2015gba,Bauer:2017fsw}. Apart from Higgs signal strength measurement, the most stringent bound on charged scalar mass comes from flavor physics when the $B\rightarrow X_s \gamma$ decay is taken into account~\cite{Amhis:2016xyh,Arbey:2017gmh}. It is found that for all types of 2HDM including type-I 2HDM, inclusive decay $b\rightarrow s \gamma$ excludes charged Higgs mass below 650 GeV for $\tan\beta<1$~ \cite{Arbey:2017gmh}.  Therefore, in the present framework, with the alignment limit $(\beta-\alpha)={\pi}/{2}$, we conservatively work with following conditions
\be
m_H=500~{\rm GeV},~m_A=m_{H^\pm}=650~{\rm GeV},\, 1\leq \tan\beta \leq 10\, .
\label{thdmlimit}
\ee

\subsubsection*{Dark Matter}
As mentioned before, the lightest neutral scalar $S_1$ or $S_2$ can serve the role of dark matter candidate in the model. We implement the model in  {\tt LanHEP-3.3.2}~\cite{Semenov:2008jy} and calculate relic density and direct detection of the dark matter using {\tt micrOMEGAs-4.3.5}~\cite{Belanger:2001fz}. We use the dark matter relic density observed by PLANCK~\cite{Aghanim:2018eyx} and constrain the model with the dark matter-nucleon scattering cross-section limits from direct search experiments XENON1T~\cite{Aprile:2018dbl,Aprile:2015uzo} and PandaX-II~\cite{Tan:2016zwf,Cui:2017nnn}.

\begin{table}[htb]
 \begin{center}
 \begin{tabular}{|c| c| c| c|c|} 
 \hline
 \hline
 $\kappa_{41}$ & $\kappa_{42}$ & $\kappa_{21}$ & $\kappa_{22}$ & {\rm DM candidate} \\ [0.5ex] 
 \hline\hline
 - & - & + & + & $S_1$  \\ 
 \hline
 - & - & - & - & $S_1$  \\
 \hline
 + & + & - & - &  $S_2$  \\
 \hline
 + & + & + & + & {\rm Lightest inert particle is charged} \\
 \hline
 \hline
 \end{tabular}
\end{center}
\caption{Combination of coupling parameters considered in the present work for the choice $0\leq \sin\gamma 
\leq \frac{1}{\sqrt{2}}$.}
\label{tab:2}
\end{table} 

Apart from 2HDM parameters (as in Eq.~(\ref{thdmlimit})), there are dark sector parameters in the model. We 
consider the following parameters to be independent parameters which constrain the  dark sector in the present 
formalism which are given as 
\be
 m_{S_2},\, \sin\gamma,\, \kappa_{1i},\, \kappa_{2i},\,\kappa_{3i},\, \kappa_{4i}\,~(i=1,2).
\label{parameters}
\ee
With the above choice of parameters the mixing between charged scalar particles ($\delta$) becomes a 
dependent parameter. 
Masses of other scalars in dark sector $S_1,~S_{1,2}^{\pm}$ and $D^{\pm\pm}$ are also obtained using these free parameters.
Moreover, we consider the case where the mixing between the neutral scalar particles 
satisfy the following condition $0\leq \sin\gamma \leq \frac{1}{\sqrt{2}}$. For simplicity we also work with the 
following condition $\kappa_{11}=\kappa_{12}$, $\kappa_{21}=\kappa_{22}$, $\kappa_{31}=\kappa_{32}$ 
and $\kappa_{41}=\kappa_{42}~(\neq 0)$. With the above mentioned choice of scalar mixing, we consider a parameter 
space depending on the choice of $\kappa_{4i}$ and $\kappa_{2i}\,(i=1,2)$ in order to obtain a viable DM 
candidate in our model. We tabulate how the choice of couplings determines whether DM is neutral or charged 
as shown in Table~\ref{tab:2}. It is to be noted that our conclusion in Table~\ref{tab:2} is independent of the 
choice of couplings $\kappa_{1i}$ and $\kappa_{3i}\,(i=1,2)$. Therefore, in the present work, we consider 
only six relevant parameters to explore DM phenomenology, are given as
\be
m_{S_2},\, \sin\gamma,\,\kappa_{2i},\, \kappa_{4i}\, ~(i=1,2).
\ee 
As observed in Table~\ref{tab:2}, we exclude the case with $\kappa_{21,22}>0$ and 
$\kappa_{41,42}>0$ values. Therefore, we are left with two cases, I) neutral DM candidate is represented by 
$S_1$ and II)  DM is represented by $S_2$. We will discuss both the cases separately in
details in this section.

\begin{figure}[!ht]
\centering
\subfigure[]{
\includegraphics[height=6 cm, width=7 cm,angle=0]{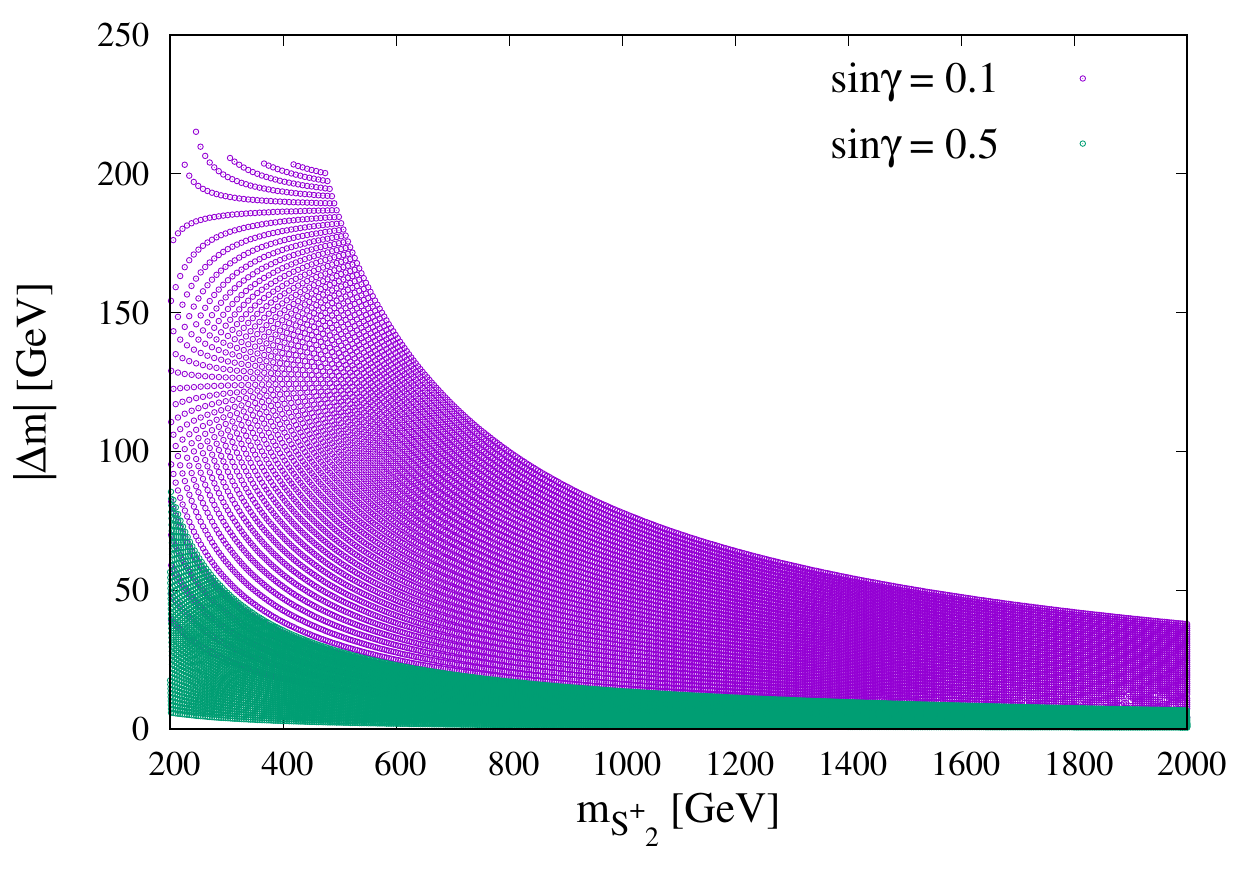}}
\subfigure []{
\includegraphics[height=6 cm, width=7 cm,angle=0]{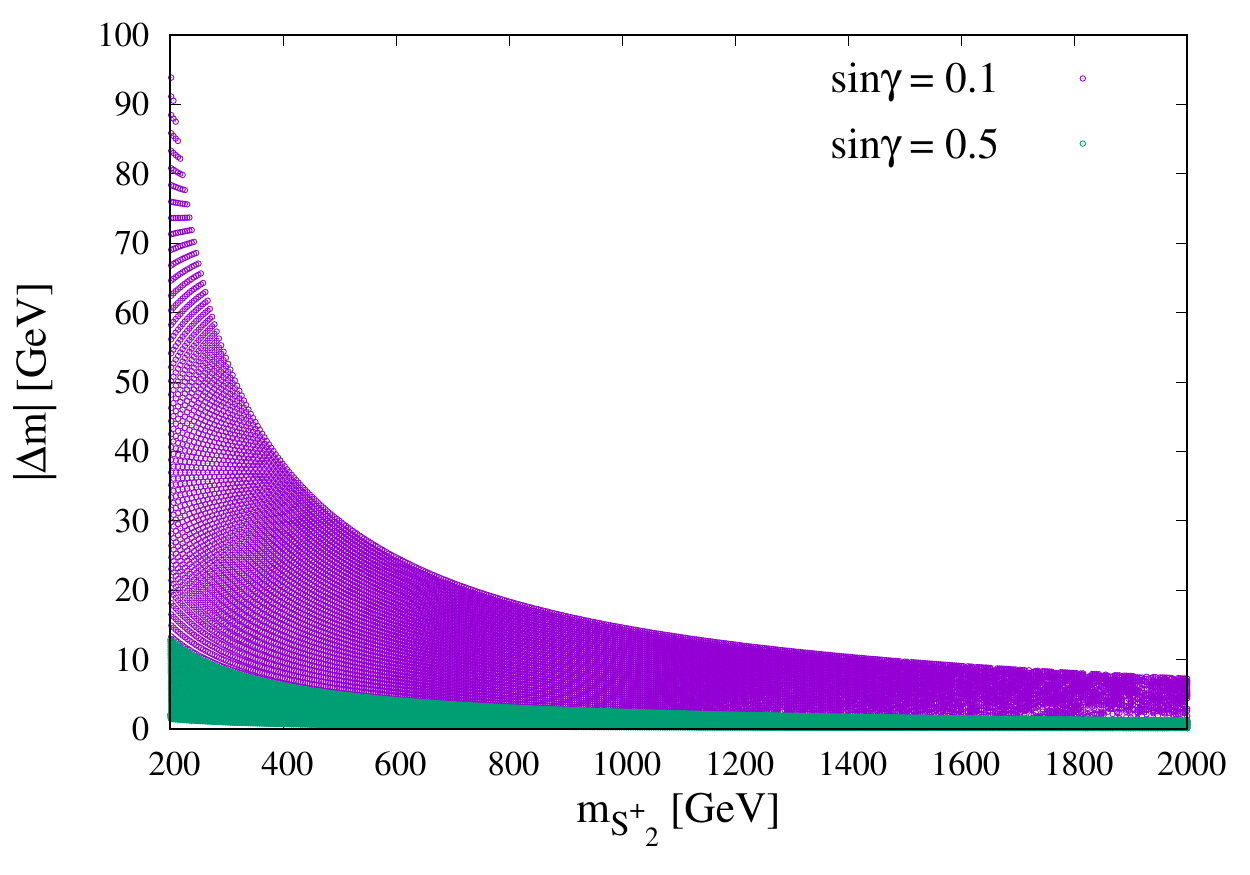}}
\caption{Allowed range of mass splitting plotted against $m_{S_2^+}$ for different values of neutral
scalar mixing angles with $\tan\beta=1$ (left panel) and $\tan\beta=10$ (right panel).}
\label{Tpara}
\end{figure}

Before we discuss the DM phenomenology, we first consider the model parameter space in agreement with 
$T$ parameter results. We consider a set of parameters following Table~\ref{tab:2}, a) $\kappa_{1i}=\kappa_{2i}=
\kappa_{3i}=0.05$ with $\kappa_{4i}$ varying in the range of $[-0.5, -0.05]$ and b) $\kappa_{1i}=-\kappa_{2i}=\kappa_{3i}
=0.05$ with $\kappa_{4i}$ varying in $[0.05, 0.5]$ ($i=1,2$). The above choice of parameters corresponds to 
first (third) row of Table~\ref{tab:2} resulting $S_1~(S_2)$ as the DM candidate. Since $T$ parameter depends 
on the mass splitting, we observe that it is independent of the choice of sign of couplings mentioned in 
Table~\ref{tab:2} and thus we ignore the second row of Table~\ref{tab:2}. We further consider two different values
of neutral scalar mixing angle $\sin\gamma=0.1,~0.5$ and vary mass $m_{S_2}$ within the range 200 GeV to 2 TeV. With 
the above choice of parameters, one can easily derive masses of other dark sector particles and also the 
mixing angle $\delta$ between charged particles. Using the bound on $T$ parameter, in Fig.~\ref{Tpara}(a) we
plot the allowed range of mass splitting $|\Delta m|=|m_{S^+_2}-m_{S^+_1}|$ against $m_{S_2^+}$ derived 
following Eq.~(\ref{Tgm}) for $\tan\beta=1$.  From Fig.~\ref{Tpara}(a), it can be easily stated that, larger values of 
mass splitting are allowed for smaller values of mixing angle. It is to be noted that for small mixing 
$\sin\gamma=0.1$, triplet scalars are almost decoupled from each other which allows larger mass splitting values
compared to the case with larger mixing $\sin\gamma=0.5$. One can also notice that as the mass $m_{S_2^+}$ is 
increased, mass difference of charged particles reduces and tends to become degenerate. 
Decrease in mass splitting enhances the possibility of 
co-annihilation of dark matter particles which can contribute to dark matter relic abundance. In, Fig.~\ref{Tpara}(b), 
we calculate the allowed range of parameter space for the same set of mixing angles $\sin\gamma$ and 
couplings for $\tan\beta=10$. We observe that larger values of $\tan\beta$ constraints the model parameter 
space significantly allowing lesser values of mass splitting in agreement with the bound on $T$ parameter. This is
intriguing because of the fact that mass splitting between particles depends on both $v_1$ and $v_2$ vacuum 
expectation values.

\subsubsection*{Case:I,  $S_1$ dark matter}
In this section we study DM phenomenology for the case when DM is represented by $S_1$ candidate and 
resembles the $T^0$ candidate of $Y=0$ triplet. We work with the conditions set by Table~\ref{tab:2} for this 
purpose. 

\begin{itemize}
\item Effects of $\sin\gamma$
\end{itemize}

\begin{figure}[!ht]
\centering
\subfigure[]{
\includegraphics[height=6 cm, width=7 cm,angle=0]{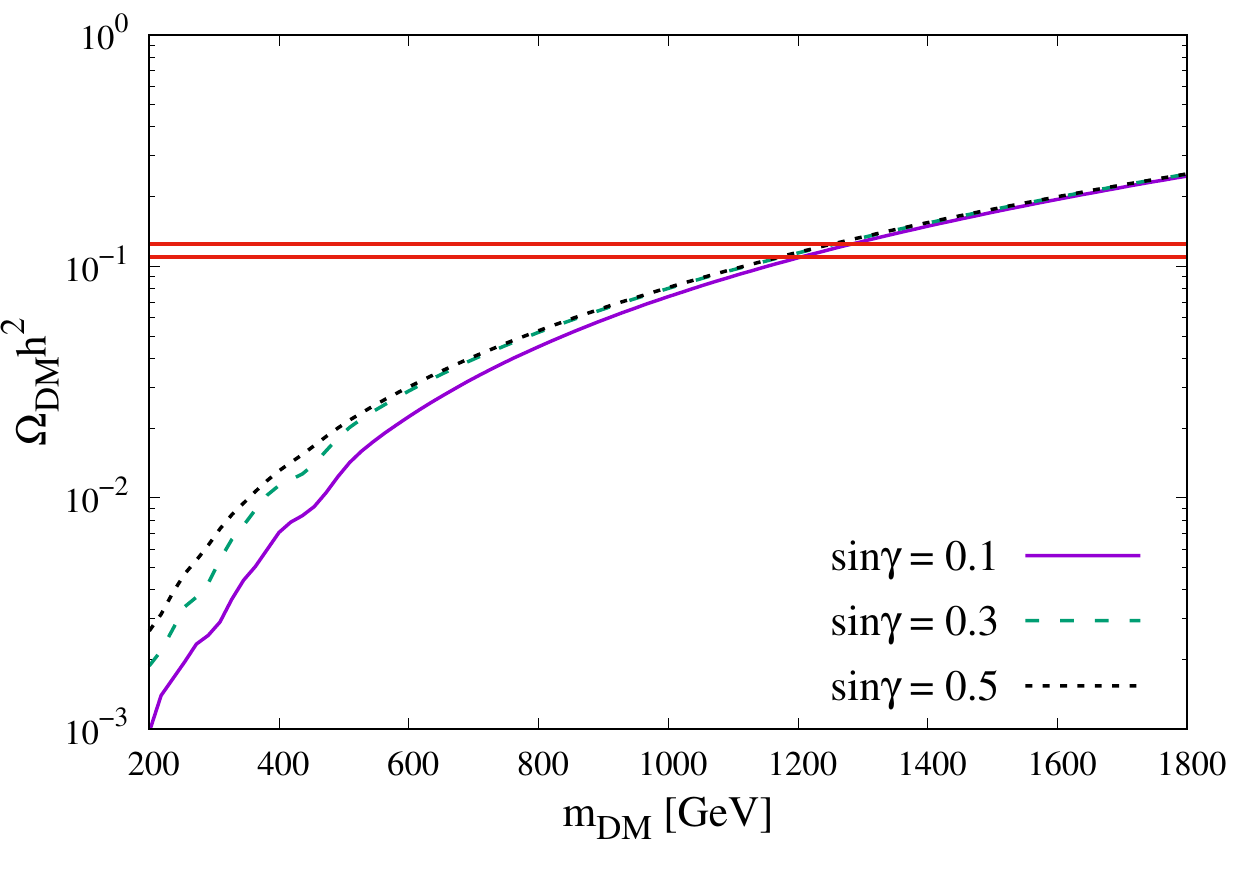}}
\subfigure []{
\includegraphics[height=6 cm, width=7 cm,angle=0]{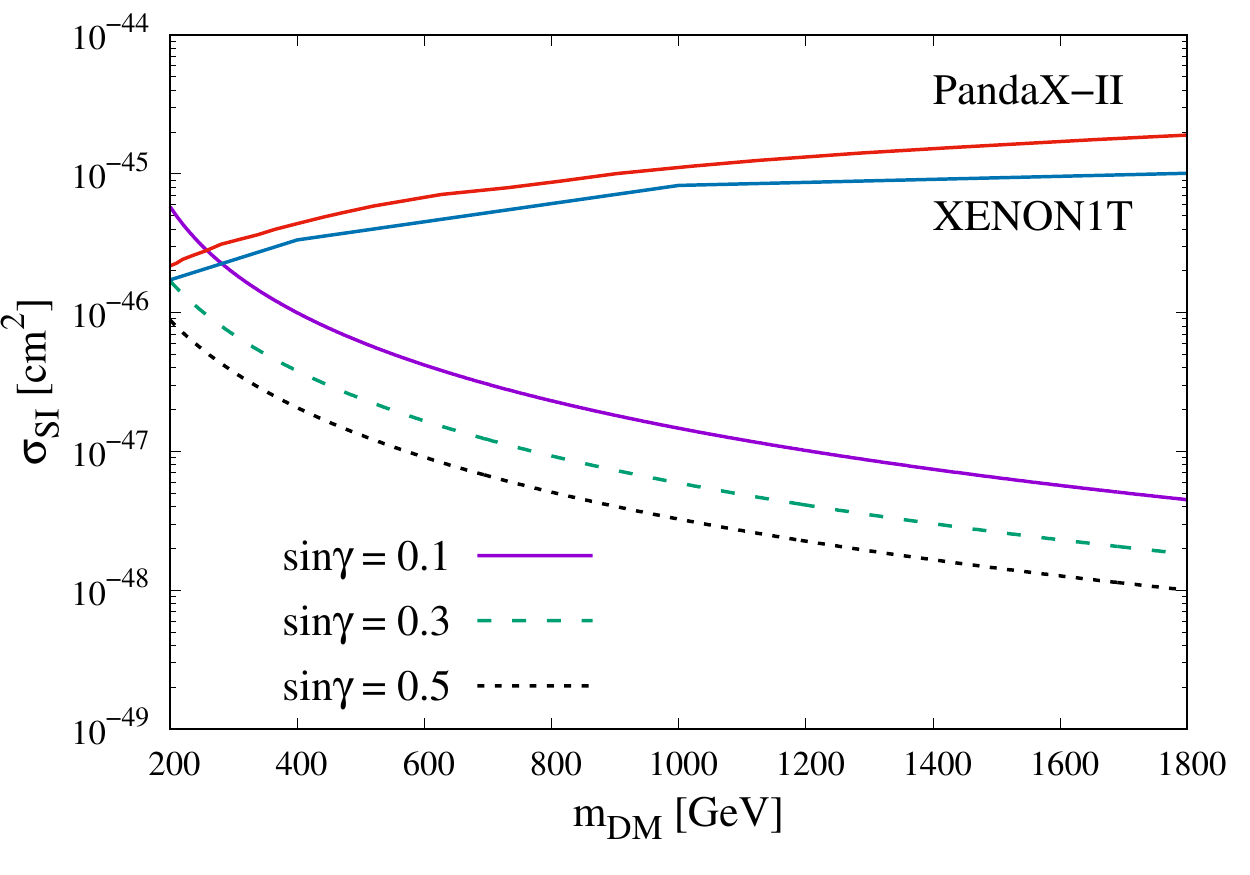}}
\caption{Left panel: Variation of DM mass with relic abundance for different values of mixing angle 
$\gamma$ (see text for details). Right panel: DM mass plotted against direct detection cross-section for 
the same set of parameters. (For interpretation of the colors in the figure(s), the
reader is referred to the web version of this article.)}
\label{fig3}
\end{figure}

In Fig.~\ref{fig3}(a), we show the variation of DM relic density with mass of DM for three different 
values of $\sin\gamma=(0.1\,,0.3\,,0.5)$. Fig.~\ref{fig3}(a) is plotted for $\tan\beta=1$, $\kappa_{1i}=\kappa_{2i}
=\kappa_{3i}=0.05$ and $\kappa_{4i}=-0.05$ where $i=1,2$. In Fig.~\ref{fig3}(b) the variation of 
dark matter direct detection against DM mass is depicted for same set of parameters and compares with the 
experimental bounds on DM-nucleon scattering cross-section with PandaX-II and XENON1T. It can be 
observed that changes in the mixing angle $\sin\gamma$ does not affect DM relic 
density very much and a 1.2 TeV dark matter candidate is in agreement with the relic density bound from 
PLANCK (red horizontal lines in left panel of Fig.~\ref{fig3}). However, with increase in the mixing angle, the 
DM direct detection cross-section tends to decreases considerably. Therefore, large mixing angle 
$\sin\gamma$ is favorable for a viable DM candidate in the present framework.   

\begin{itemize}
\item Effects of $\tan\beta$
\end{itemize}

\begin{figure}[!ht]
\centering
\subfigure[]{
\includegraphics[height=6 cm, width=7 cm,angle=0]{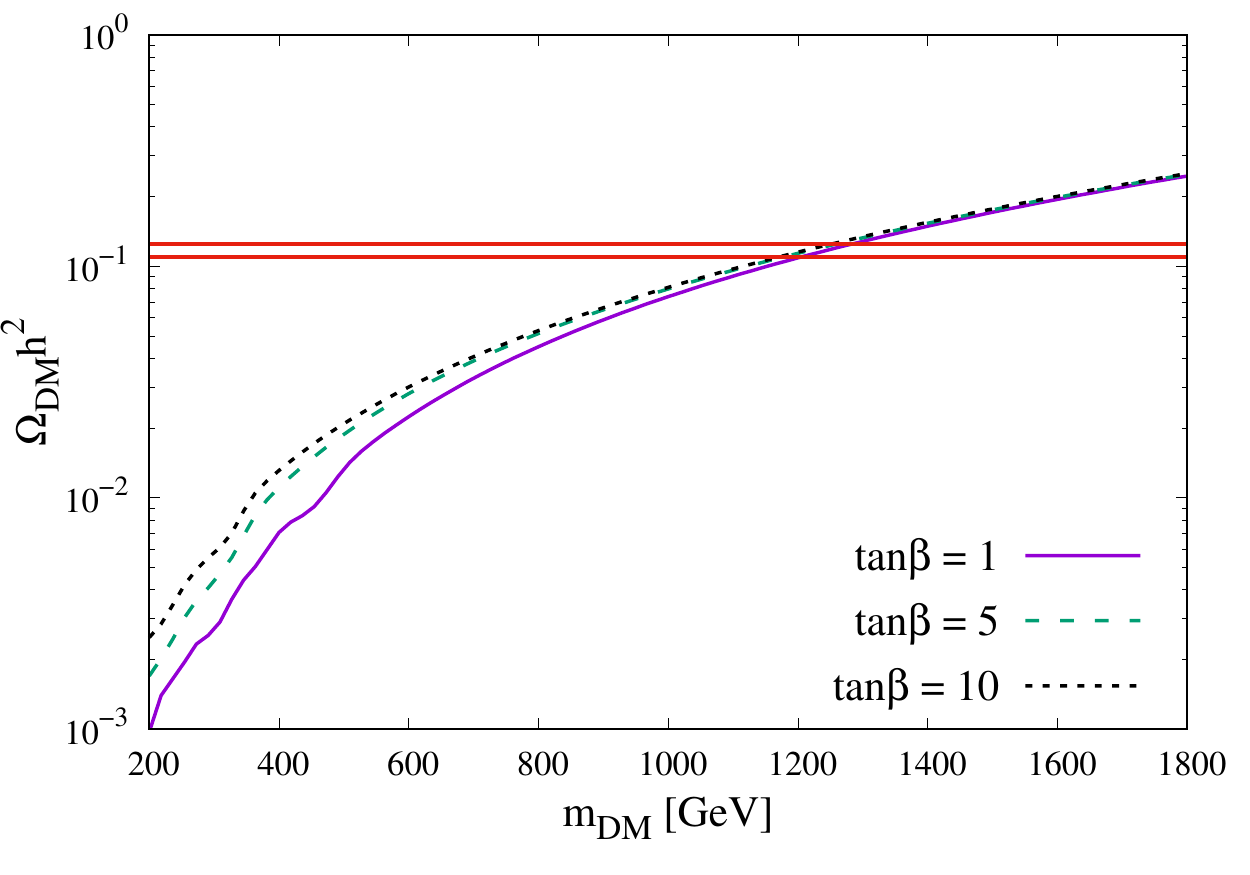}}
\subfigure []{
\includegraphics[height=6 cm, width=7 cm,angle=0]{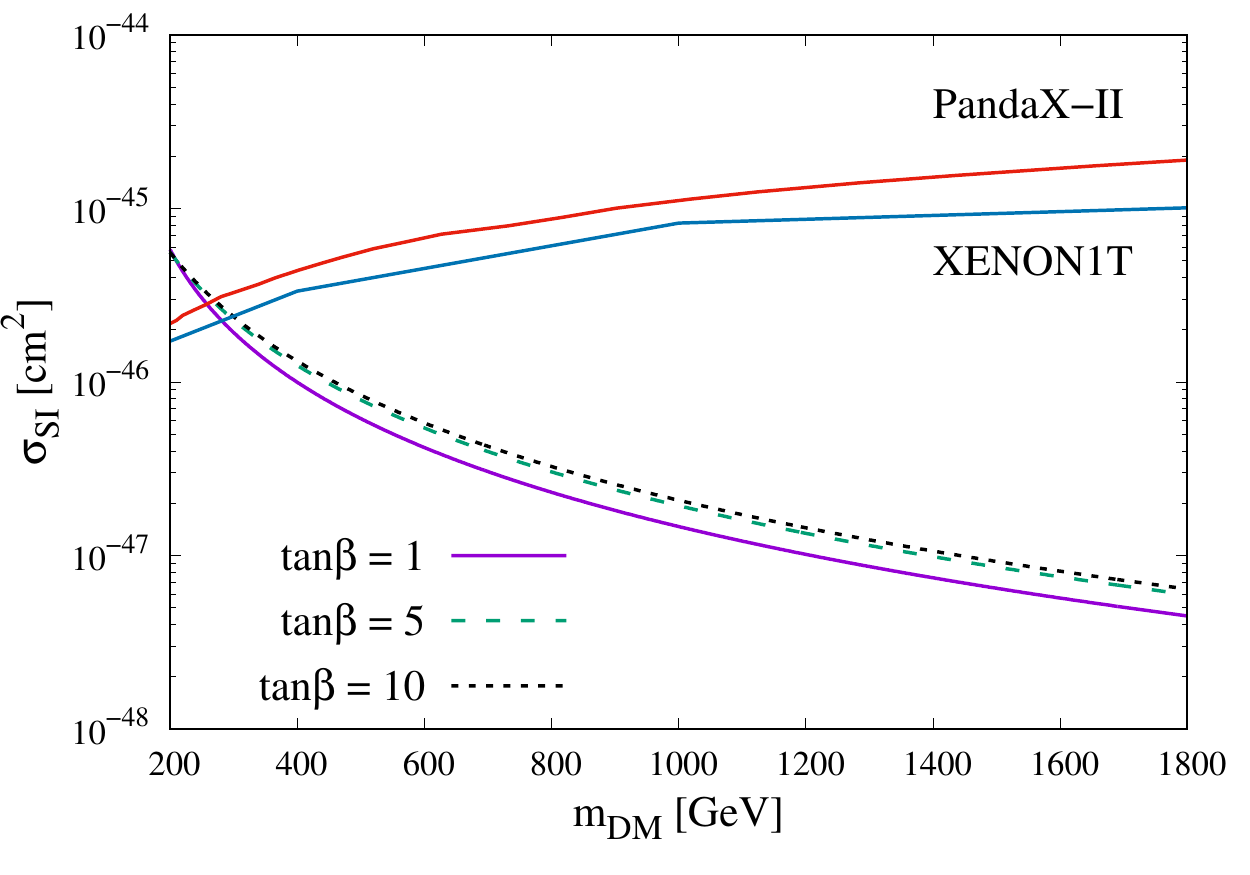}}
\caption{Left panel: DM mass vs relic density for different $\tan\beta$. Right panel: DM mass vs direct detection 
cross-section plotted with same set of $\tan\beta$ values. See text for details.}
\label{fig4}
\end{figure}

In Fig.~\ref{fig4}, we repeat the results for dark matter with same set of coupling parameters as used in 
case of Fig.~\ref{fig3} for different values of $\tan\beta=(1\,,5\,,10)$ with scalar mixing fixed at
$\sin\gamma=0.1$. We observe in Fig.~\ref{fig4}(a) that for different values of $\tan\beta$,
DM relic abundance does not change significantly with DM mass. However, DM-nucleon scattering 
cross-section increases with larger $\tan\beta$ values as seen in Fig.~\ref{fig4}(b).
\begin{itemize}
\item Effects of $\kappa_{41,42}$
\end{itemize}

\begin{figure}[!ht]
\centering
\includegraphics[height=6 cm, width=7 cm,angle=0]{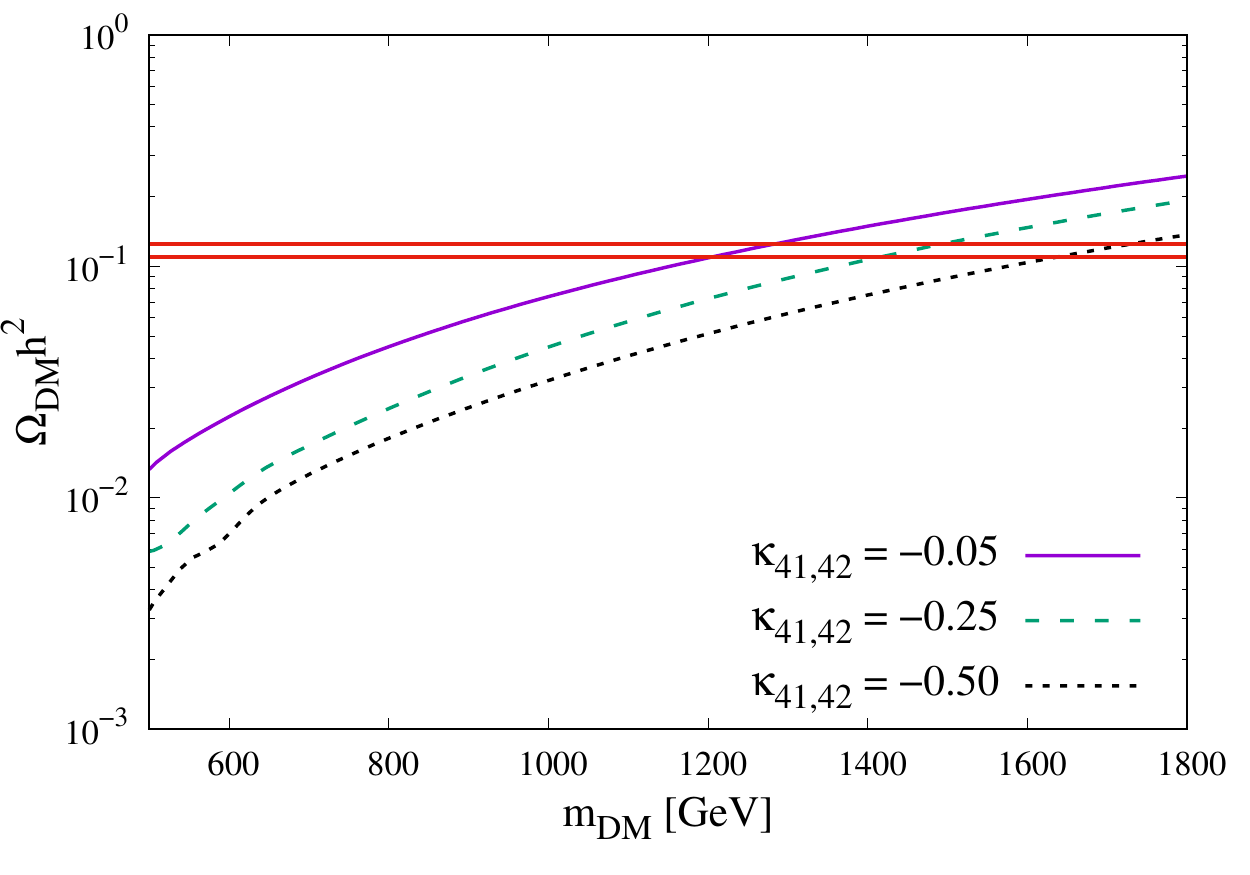}
\caption{Variation of DM relic abundance with DM mass for different values of $\kappa_{41,42}$ 
using $\sin\gamma=0.1$ and $\tan\beta=1$.}
\label{fig5}
\end{figure}

Fig.~\ref{fig5} shows the variation of DM relic density against DM mass for three different values of $\kappa_{41,42}$ for $\tan\beta=1$ and $\sin\gamma=0.1$. Other parameter are kept fixed at same values as considered in Fig.~\ref{fig3}. In Fig.~\ref{fig5} we observe that as $\kappa_{41,42}$ changes from -0.05 to -0.5, relic density plots changes significantly which allows a range of dark matter mass from 1.2 TeV to 1.7 TeV that is in agreement with PLANCK results. Increase in $|\kappa_{41,42}|$ value results in increase in mass splitting between charged and neutral eigenstates $S^+_{1,2}$ and $S_{1,2}$. With increased mass splitting, contribution from co-annihilation channels decreases which reduces DM relic density for a given mass of DM as observed in Fig.~\ref{fig5}. However, no significant change in DM direct detection is observed for changes in coupling $\kappa_{41,42}$.

\begin{itemize}
\item Effects of $\kappa_{11,12}$ and $\kappa_{31,32}$
\end{itemize}

\begin{figure}[!ht]
\centering
\includegraphics[height=6 cm, width=7 cm,angle=0]{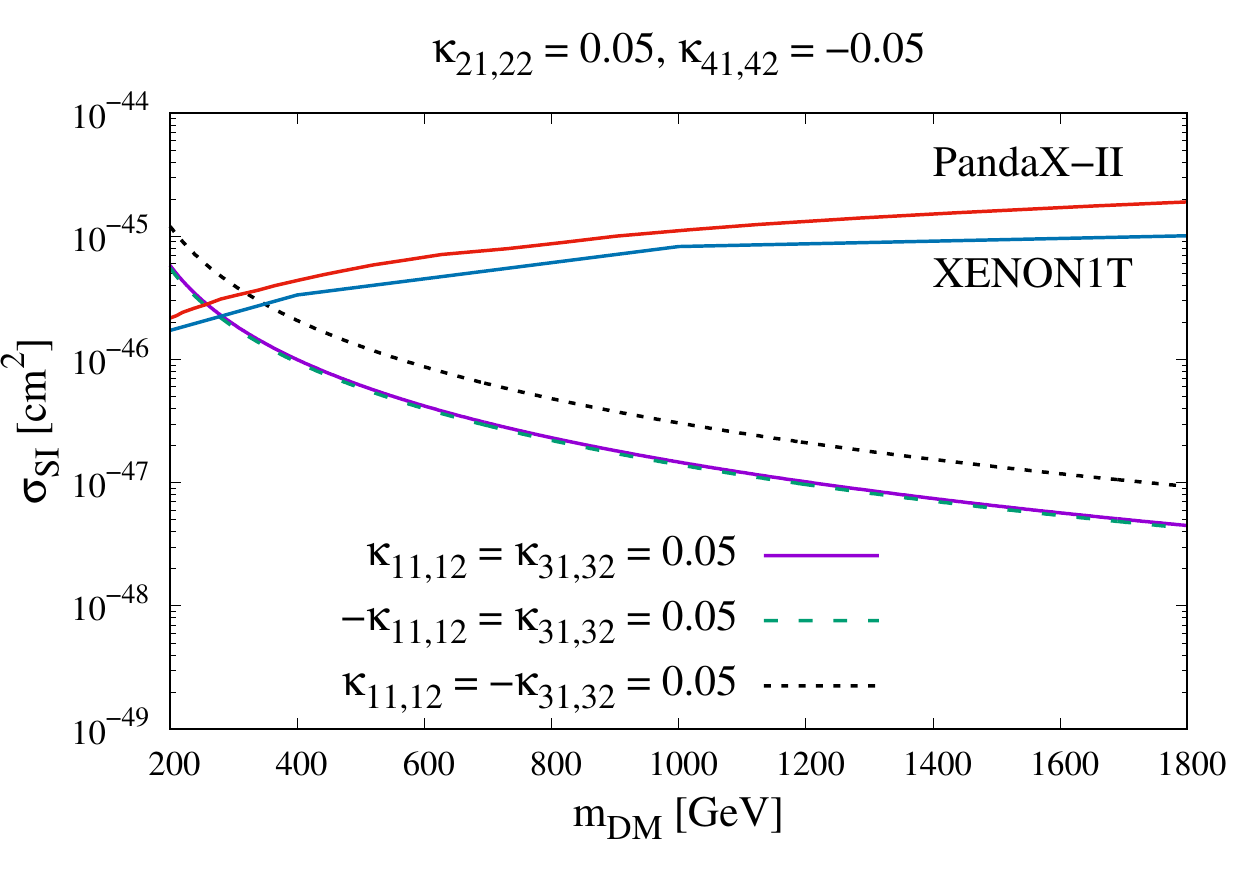}
\caption{DM mass vs $\sigma_{\rm SI}$ plots for different combinations of $\kappa_{11,12}$ and 
$\kappa_{31,32}$ using $\sin\gamma=0.1$ and $\tan\beta=1$.}
\label{fig8a}
\end{figure}

Finally, we consider changes of couplings $\kappa_{1i}$ and $\kappa_{3i}$, $i=1,2$ and study how these couplings affect dark matter phenomenology. It is found from the study that couplings $\kappa_{1i}$ and $\kappa_{3i}$ does not alter DM relic density significantly but can affect the DM direct detection cross-sections. This is obvious since, these couplings allow Higgs portal interactions with SM sector particles. However, since gauge interactions and co-annihilation channels dominate over Higgs portal interactions, these couplings are less sensitive to DM relic density.   For the purpose of demonstration, we consider $\tan\beta=1$, $\sin\gamma=0.1$ and couplings $\kappa_{21,22},~\kappa_{41,42}$ etc. used in previous set of plots shown in Fig.~\ref{fig3}-\ref{fig4}.  From Fig.~\ref{fig8a}, we observe that changing value of $\kappa_{11,12}=-0.05$ from $\kappa_{11,12}=0.05$, does not affect direct detection results significantly. However, for the case $\kappa_{31,32}=-0.05$ an increase in DM direct detection is observed when compared with the case $\kappa_{31,32}=0.05$. Therefore, we can conclude that $\kappa_{31,32}<0$ enhances DM-nucleon scattering cross-section in the present framework.
 
It is to be noted that for DM candidate ($S_1$) there exists another region of parameter space with positive values of $\kappa_{21,22}$ as mentioned earlier in Table~\ref{tab:2}. We have found that the results for the region with positive $\kappa_{21,22}$ also follow similar behaviour as obtained in Figs.~\ref{fig3}-\ref{fig8a}. Hence, DM phenomenology for the case with $\kappa_{21,22}>0$ are not shown in this work. 

\subsubsection*{Case:II, $S_2$ dark matter}
In the previous section, we have presented the results for DM dominated by $S_1$. In this section we consider the case when DM is $S_2$ dominated which resemble the neutral candidate of $Y=1$ triplet. For this purpose, we follow the conditions mentioned in Table~\ref{tab:2}.

\begin{itemize}
\item Effects of $\sin\gamma$
\end{itemize}

\begin{figure}[!ht]
\centering
\subfigure[]{
\includegraphics[height=6 cm, width=7 cm,angle=0]{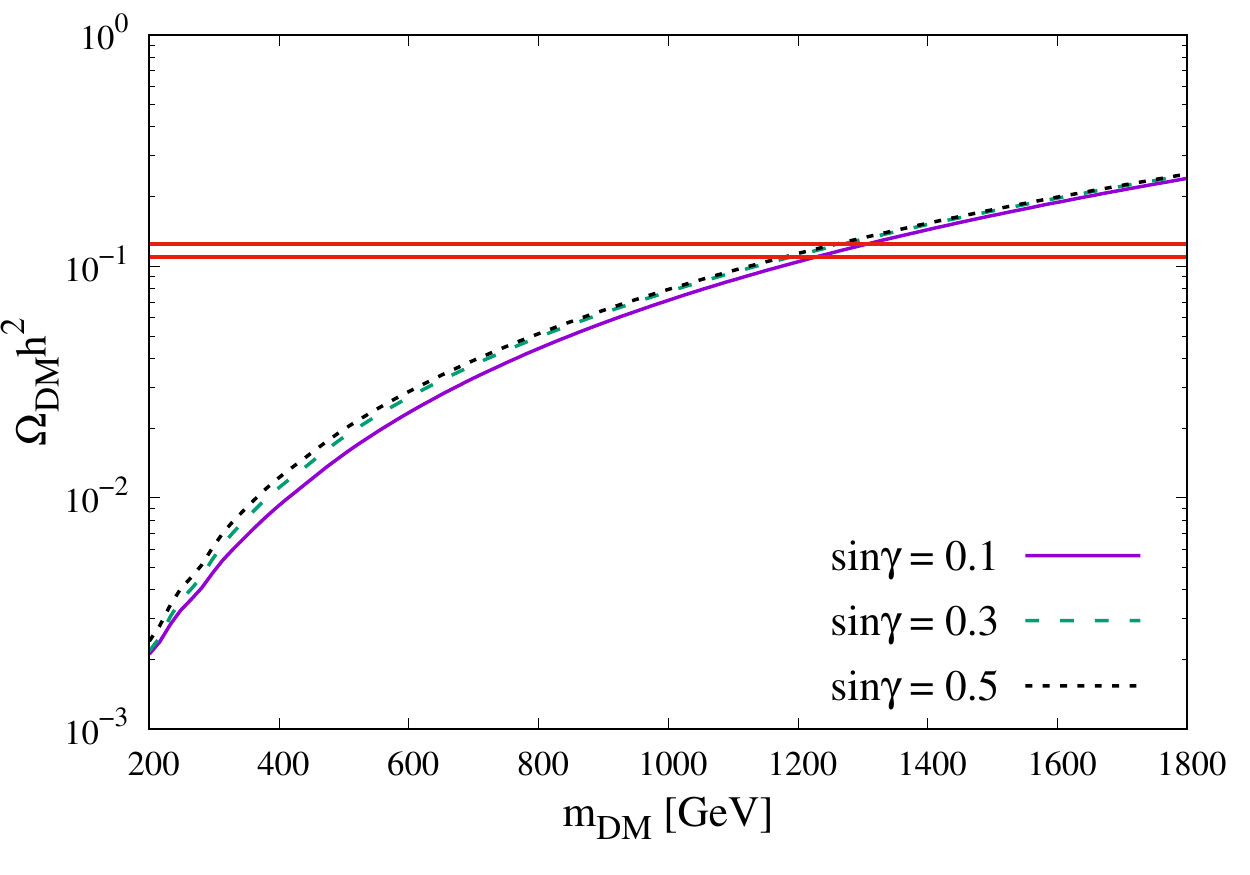}}
\subfigure []{
\includegraphics[height=6 cm, width=7 cm,angle=0]{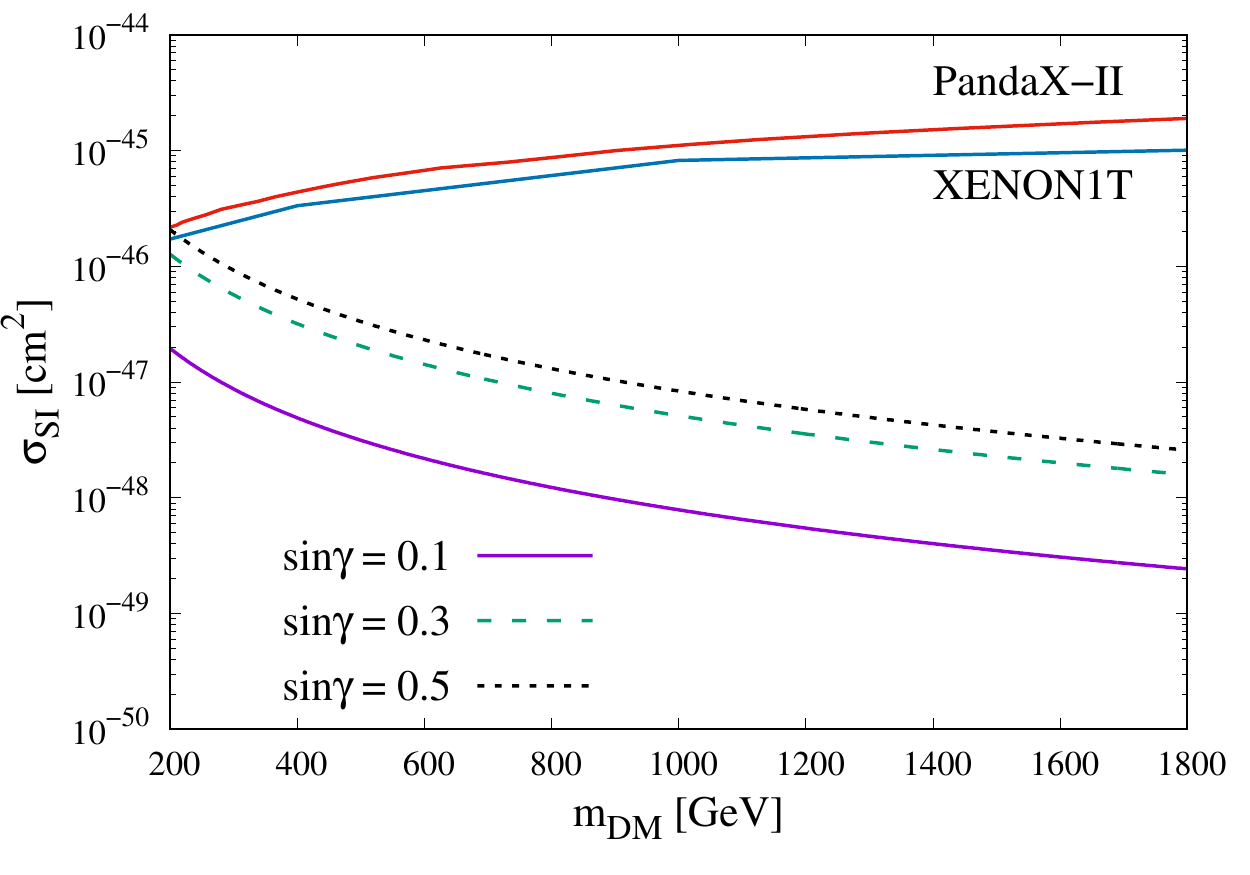}}
\caption{Left panel: Variation of DM mass with relic abundance for different $\sin\gamma$. See text for details. 
Right panel: DM mass plotted against $\sigma_{\rm SI}$ for the same set of parameters.}
\label{fig6}
\end{figure}

Similar to the previous section, we now repeat our study for the case considering DM candidate is $S_2$ with positive values of $\kappa_{41,42}$ following Table~\ref{tab:2}. In Fig.~\ref{fig6}(a)-(b), we show the variation of DM relic density and direct detection cross-section against DM mass for three different values of $\sin\gamma$ plotted with set of coupling parameters $\kappa_{11,12}=\kappa_{31,32}=\kappa_{41,42}=0.05$ and $\kappa_{21,22}=-0.05$ using $\tan\beta=1$. We observe that, even DM candidate is  $S_2$-like, with different $\sin\gamma$ values, DM mass versus relic density plots does not undergo any significant changes and 1.2 TeV dark matter satisfies DM relic density. However, significant changes in DM mass versus DM-nucleon scattering cross-section plot can be observed with changes in $\sin\gamma$ values as depicted in Fig.~\ref{fig6}(b).  It is clearly shown that with increasing $\sin\gamma$, the DM direct detection cross-section gains large enhancement. This is quite different when compared with the case of $S_1$ DM case presented in Fig.~\ref{fig3}(b). 
\begin{itemize}
\item Effects of $\tan\beta$
\end{itemize}

\begin{figure}[!ht]
\centering
\subfigure[]{
\includegraphics[height=6 cm, width=7 cm,angle=0]{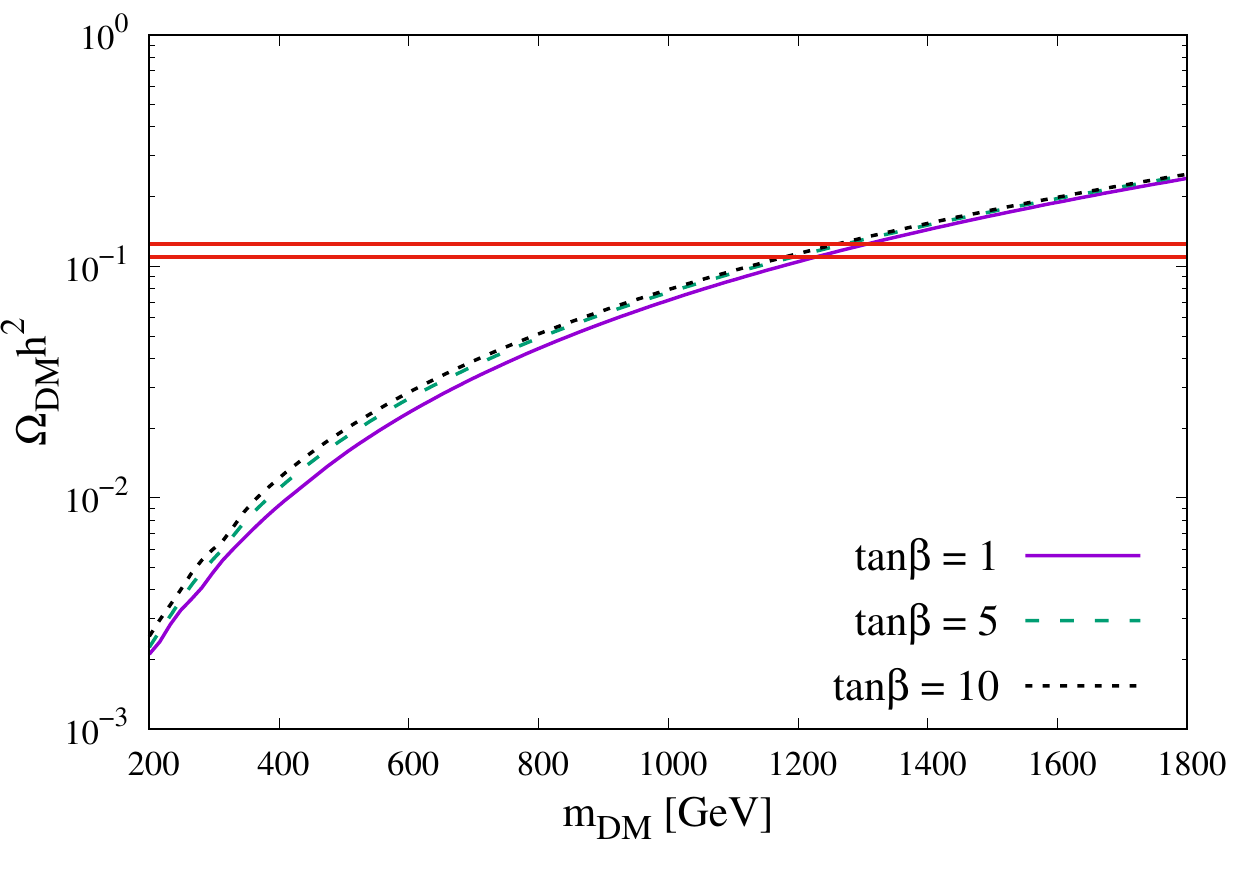}}
\subfigure []{
\includegraphics[height=6 cm, width=7 cm,angle=0]{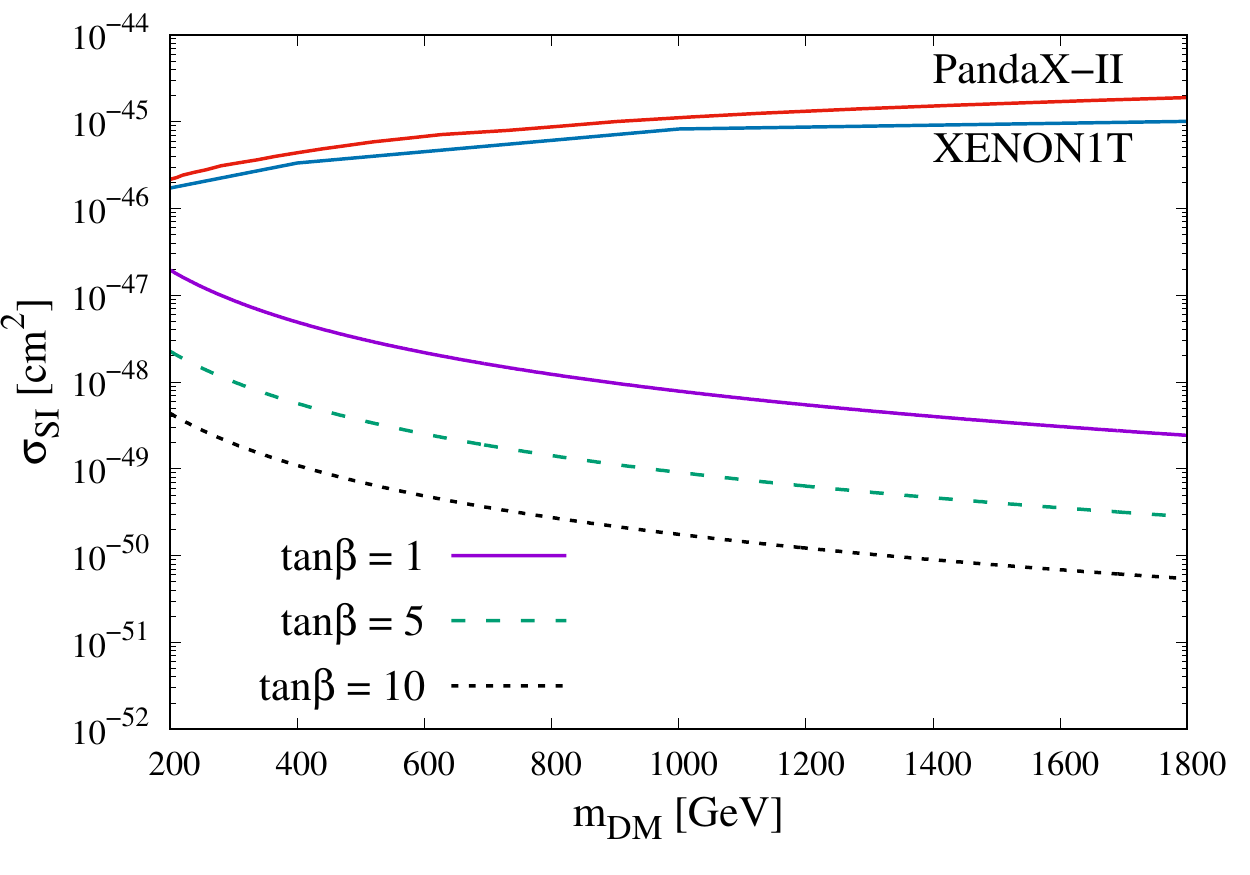}}
\caption{Left panel: Dark matter mass vs relic density for different values of $\tan\beta$. See text for details. 
Right panel: DM mass against DM nucleon scattering cross-section for same set of $\tan\beta$ and other 
parameters.}
\label{fig7}
\end{figure}

We now repeat the study with three different $\tan\beta=1,~5,~10$ values for $\sin\gamma=0.1$, keeping other 
parameters fixed as taken in previous case (Fig.~\ref{fig6}), and present our findings in Fig.~\ref{fig7}. 
From Fig.~\ref{fig7}(a), we conclude that changes in 
$\tan\beta$ is less significant to DM relic density. However, Fig.~\ref{fig7}(b) predicts that for larger $\tan\beta$, 
DM-nucleon cross-section tends to decrease. 

\begin{itemize}
\item Effects of $\kappa_{41,42}$
\end{itemize}
\begin{figure}[!htb]
\centering
\includegraphics[height=6 cm, width=7 cm, angle=0]{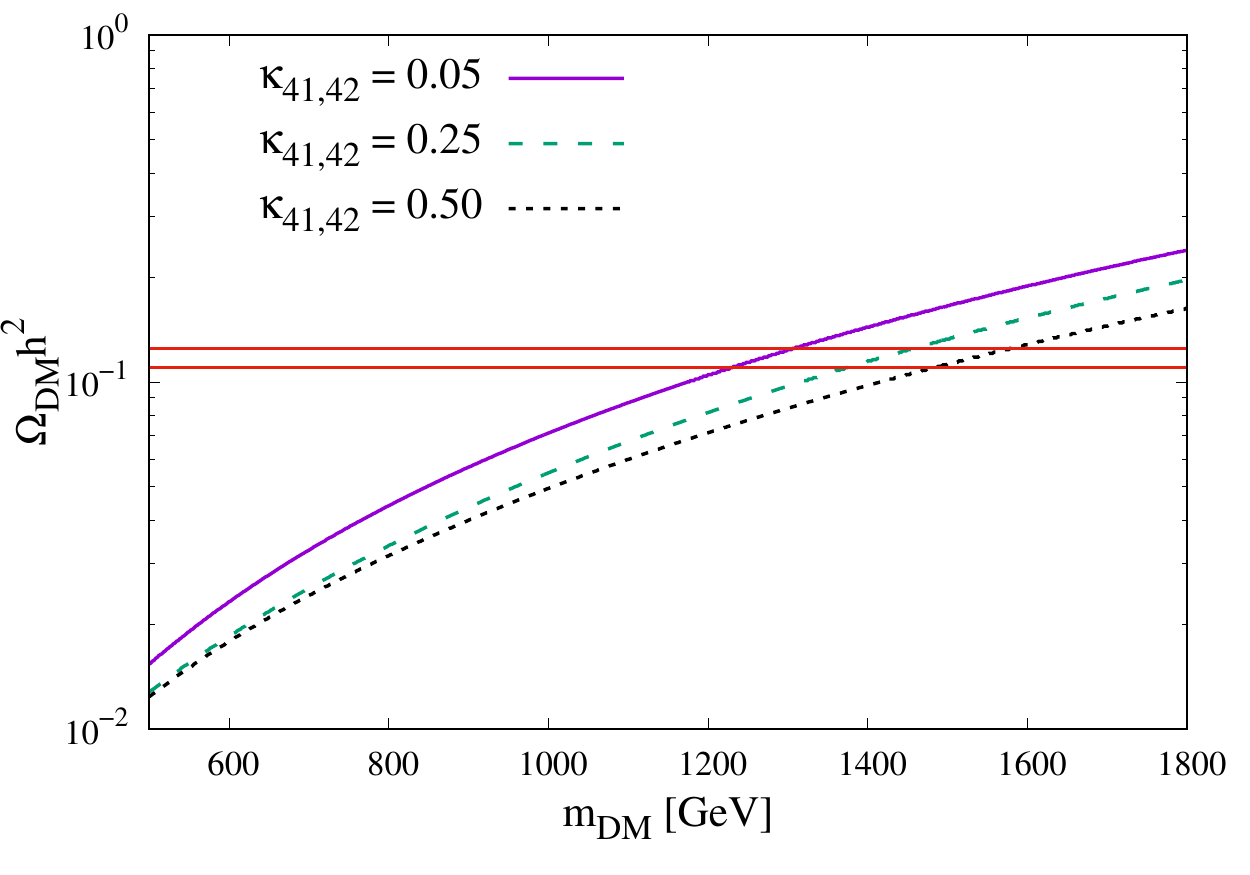}
\caption{DM mass vs relic density plots for different values of $\kappa_{41,42}$ using $\sin\gamma=0.1$
and $\tan\beta=1$.}
\label{fig8}
\end{figure}
In Fig.~\ref{fig8}, we repeat our study following Fig.~\ref{fig5} by varying $\kappa_{41,42}$ from 0.05 to 0.5 for
$\sin\gamma=0.1$ and $\tan\beta=1$ keeping the rest of the parameters unchanged as considered in 
Fig.~\ref{fig6}. Similar to the case of dark matter in Fig.~\ref{fig5}, here we also observe large 
changes in DM relic density which predicts that DM can have mass around 1.2 TeV to 1.6 TeV. This can be explained
by the rise in mass splitting between dark sector particles which causes reduction in DM annihilation
cross-section happening from DM co-annihilation. Note that, for different values of $\kappa_{4i}\,(i=1-2)$, DM 
direct detection cross-section remains unchanged similar to the case of $S_1$ dark matter.

\begin{itemize}
\item Effects of $\kappa_{11,12}$ and $\kappa_{31,32}$
\end{itemize}

\begin{figure}[!ht]
\centering
\includegraphics[height=6 cm, width=7 cm,angle=0]{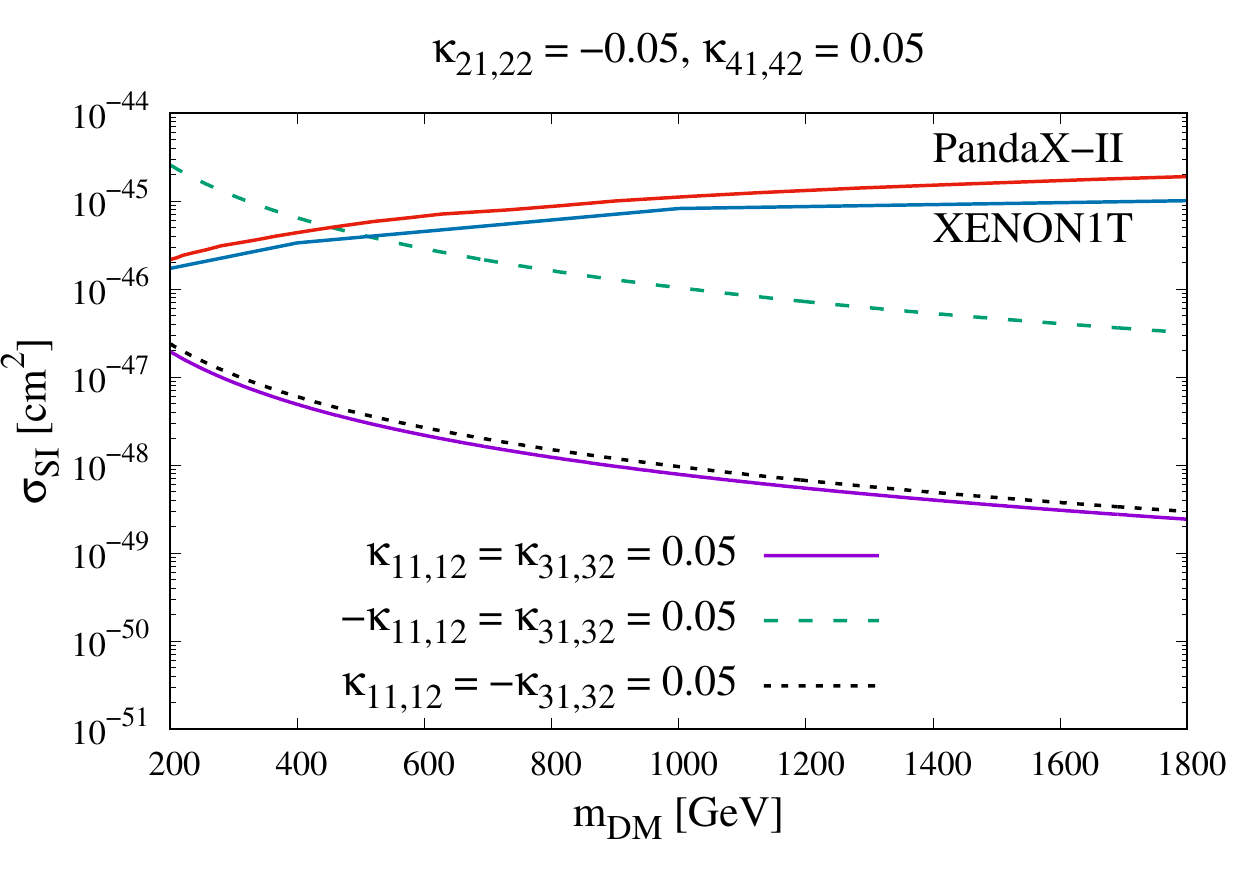}
\caption{ Variation of DM mass with $\sigma_{\rm SI}$ for different choices of $\kappa_{11,12}$ and $\kappa_{31,32}$ 
using $\sin\gamma=0.1$ and $\tan\beta=1$.}
\label{fig8b}
\end{figure}

Similar to the study of $S_1$ dark matter phenomenology performed  in Fig.~\ref{fig8a}, we investigate the effects of $\kappa_{11,12}$ and $\kappa_{31,32}$ in Fig.~\ref{fig8b} on DM-nucleon scattering cross-section. As we have  mentioned before, due to the effects of gauge annihilations and co-annihilation of dark matter candidate, these couplings are not significant to DM relic abundance. Looking into Fig.~\ref{fig8b}, we conclude that dark matter direct detection results remains almost unaffected with change in sign of $\kappa_{31,32}$ from positive to negative. However, for the case $\kappa_{11,12}=-0.05$, there is significant enhancement in DM direct detection cross-section when compared with the case $\kappa_{11,12}=0.05$, which can be probed by next generation DM direct search experiments and constrain the available model parameter space.

\begin{table}[htb]
 \begin{center}
 \begin{tabular}{|c| c| c| c|c|} 
 \hline
 \hline
 $\kappa_{41}$ & $\kappa_{42}$ & $\kappa_{21}$ & $\kappa_{22}$ & {\rm DM candidate} \\ [0.5ex] 
 \hline\hline
 + & + & + & + & $S_1$  \\ 
 \hline
 + & + & - & - & $S_1$  \\
 \hline
 - & - & - & - &  $S_2$  \\
 \hline
 - & - & + & + & {\rm Lightest inert particle is charged} \\
 \hline
 \hline
 \end{tabular}
\end{center}
\caption{Combination of different coupling parameters for the choice $-\frac{1}{\sqrt{2}}\leq \sin\gamma \leq 0$.}
\label{tab:3}
\end{table} 

It is to be noted that apart from the region of parameter space considered in Table~\ref{tab:2}, there also exists another region of available parameter space as given in Table~\ref{tab:3} for $-\frac{1}{\sqrt{2}}\le \sin\gamma \le 0$. However, the DM phenomenology does not alter significantly for negative values of neutral scalar mixing $\sin\gamma$ and results are similar to cases discussed in Figs.~\ref{fig3}-\ref{fig8}. Therefore, we will not discuss the combination of parameter space mentioned in Table~\ref{tab:3}. 

Therefore, DM phenomenology of 2HDM extension of inert GM model suggests that a viable TeV scale dark matter can achieved within the framework. Since the masses of dark sector particles are around few TeVs, their contribution to the process $h\rightarrow \gamma \gamma$ is negligible compared to the contribution from $W^\pm,~H^\pm$ and fermion (top and bottom quarks) and thus can be ignored safely. In other words, conditions described in Eq.~(\ref{thdmlimit}) for type-I 2HDM remains unaltered.  

\subsubsection*{Lepton flavor violation}

\begin{figure}[!ht]
\centering
\begin{minipage}{0.48\linewidth}
\centering
\begin{tikzpicture}[line width=1.5 pt, scale=1.3,  >=latex]
	\draw[fermion] (-2,0)--(-1,0);
	\draw[fermion] (-1,0)--(1,0);
	\draw[scalarbar] (-1,0) arc (180:90:1);
	\draw[scalar] (1,0) arc (0:90:1);
\begin{scope}[shift={(90:1)}]
    \draw[vector] (0,0)--(40:1.5);
    \node at (30:1.5) {$\gamma$};
\end{scope}
	\draw[fermion] (1,0)--(2,0);
	\node at (-2,-0.25) {$l_i^-$};
	\node at (0,-0.25) {$\Sigma^0$};
	\node at (2,-0.25) {$l_j^-$};
	\node at (40:1.4) {$S_{1,2}^+$};
	\node at (140:1.4) {$S_{1,2}^+$};
\end{tikzpicture}
\end{minipage}
\begin{minipage}{0.48\linewidth}
\centering
\begin{tikzpicture}[line width=1.5 pt, scale=1.3,  >=latex]
	\draw[fermion] (-2,0)--(-1,0);
	\draw[fermion] (-1,0)--(0,0);
	\draw[fermion] (0,0)--(1,0);
	\draw[scalarnoarrow] (-1,0) arc (180:0:1);
    \draw[vector] (0,0)--(315:1.5);
    \node at (325:1.5) {$\gamma$};
	\draw[fermion] (1,0)--(2,0);
	\node at (-2,-0.25) {$l_i^-$};
	\node at (-0.5,-0.25) {$\Sigma^-$};
	\node at (0.7,-0.25) {$\Sigma^-$};
	\node at (2,-0.25) {$l_j^-$};
	\node at (90:1.4) {$S_{1,2}$};
\end{tikzpicture}
\end{minipage}
\begin{minipage}{0.48\linewidth}
\centering
\begin{tikzpicture}[line width=1.5 pt, scale=1.3,  >=latex]
	\draw[fermion] (-2,0)--(-1,0);
	\draw[fermionbar] (-1,0)--(1,0);
	\draw[scalarbar] (-1,0) arc (180:90:1);
	\draw[scalar] (1,0) arc (0:90:1);
\begin{scope}[shift={(90:1)}]
    \draw[vector] (0,0)--(40:1.5);
    \node at (30:1.5) {$\gamma$};
\end{scope}
	\draw[fermion] (1,0)--(2,0);
	\node at (-2,-0.25) {$l_i^-$};
	\node at (0,-0.25) {$\Sigma^-$};
	\node at (2,-0.25) {$l_j^-$};
	\node at (40:1.4) {$D^{++}$};
	\node at (140:1.4) {$D^{++}$};
\end{tikzpicture}
\end{minipage}
\begin{minipage}{0.48\linewidth}
\centering
\begin{tikzpicture}[line width=1.5 pt, scale=1.3,  >=latex]
	\draw[fermion] (-2,0)--(-1,0);
	\draw[fermionbar] (-1,0)--(0,0);
	\draw[fermionbar] (0,0)--(1,0);
	\draw[scalarbar] (-1,0) arc (180:0:1);
    \draw[vector] (0,0)--(315:1.5);
    \node at (325:1.5) {$\gamma$};
	\draw[fermion] (1,0)--(2,0);
	\node at (-2,-0.25) {$l_i^-$};
	\node at (-0.5,-0.25) {$\Sigma^-$};
	\node at (0.7,-0.25) {$\Sigma^-$};
	\node at (2,-0.25) {$l_j^-$};
	\node at (90:1.4) {$D^{++}$};
\end{tikzpicture}
\end{minipage}
\caption{$l_i \to l_j \gamma$ diagrams at the one-loop level.}
\label{figlfv}
\end{figure}
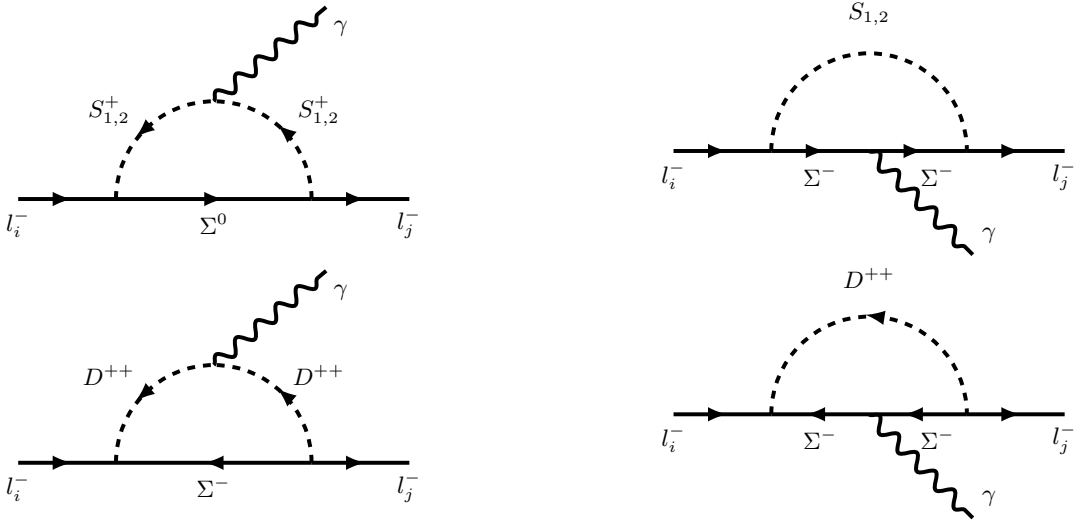
In the present model, Yukawa interactions mentioned in the Eq.~(\ref{int}) will result in charged lepton 
flavor violation. In Fig.~\ref{figlfv}, we show possible diagrams that contribute to flavor violating decay. The Feynman amplitude for the process $l_i\rightarrow l_j \gamma$ is given as 
\begin{equation}\label{muAB}
 \mathcal{M}(l_i \to l_j \gamma)=\epsilon^{\mu}\bar{u}_{l_j}(p-q)[iq^{\nu}\sigma_{\mu \nu}( F + G 
 \gamma_5)]u_{l_i}(p)\, \, .
\end{equation}
where the form factors $F$ and $G$ are expressed as
\begin{equation}
\begin{split}
F=G= &\sum_{k=1}^3 \frac{\lambda_{ik}\lambda_{jk} }{64 \pi^2 M_{\Sigma_k}^2}(m_{l_i} + m_{l_j})^2
\left(\cos^2 \delta F_3^{(a)} +\sin^2\delta F_4^{(a)} -\frac{1}{2} \cos^2 \gamma \ F_1^{(b)} -\frac{1}{2} \sin^2 
\gamma\  F_2^{(b)}\right) \\
&+\sum_{k=1}^3 \frac{y_{ik}y_{jk} }{64 \pi^2 M_{\Sigma_k}^2}(m_{l_i} + m_{l_j})^2 \left(\sin^2 \delta F_3^{(a)} +
\cos^2\delta F_4^{(a)} + F_5^{(a)} - 2 F_5^{(b)}\right)\\
&+ \sum_{k=1}^3 \frac{\lambda_{ik}y_{jk} +\lambda_{jk}y_{ik} }{64 \pi^2 M_{\Sigma_k}^2}(m_{l_i} + 
m_{l_j})^2\sin \delta \cos \delta\left(F_4^{(a)} - F_3^{(a)}\right)\,,
\end{split}
\label{lfveq1}
\end{equation}
where,
\begin{equation}
\begin{split}
  F_i^{(a)} =&\frac{1}{6(1-x_i)^4}\left(2+3x_i-6x_i^2+x_i^3+6x_i\ln x_i\right) \, ,\\
  F_i^{(b)} =&\frac{1}{6(1-x_i)^4}\left(1-6x_i-3x_i^2-2x_i^3-6x_i^2\ln x_i\right)\, ,\\
x_1=\frac{m_{S_1}^2}{M_{\Sigma}^2}\,, \qquad & x_2=\frac{m_{S_2}^2}{M_{\Sigma}^2}\,, \qquad 
x_3=\frac{m_{S_1^+}^2}{M_{\Sigma}^2}\,, \qquad x_4=\frac{m_{S_2^+}^2}{M_{\Sigma}^2}\,, \qquad
x_5=\frac{m_{D^{++}}^2}{M_{\Sigma}^2}\,.
\end{split}
\label{lfveq2}
\end{equation}
In Eq.~(\ref{lfveq2}), we have dropped the indices of $M_{\Sigma_k}$ for simplicity. It is to be noted that for a vector fermion doublet $M_{\Sigma_k^0}=M_{\Sigma_k^+}=M_{\Sigma_k}$. Stringent bound on flavor violating process $\mu \rightarrow e \gamma$  is  obtained from MEG experiment~\cite{TheMEG:2016wtm}. Upper limit on the decay branching ratio for $\mu \rightarrow e \gamma$ decay is ${\mathcal B}(\mu \rightarrow e \gamma) < 4.2 \times 10^{-13}$. The decay rate for the process $\mu \rightarrow e \gamma$ in the present model is given by
\begin{equation}
\Gamma(\mu \to e \gamma)=\frac{m_{\mu}}{8\pi}(|F|^2 + |G|^2)\, .
\end{equation}

\begin{figure}[t]
  \centering
  \subfigure[]{
\includegraphics[height=6 cm, width=7 cm,angle=0]{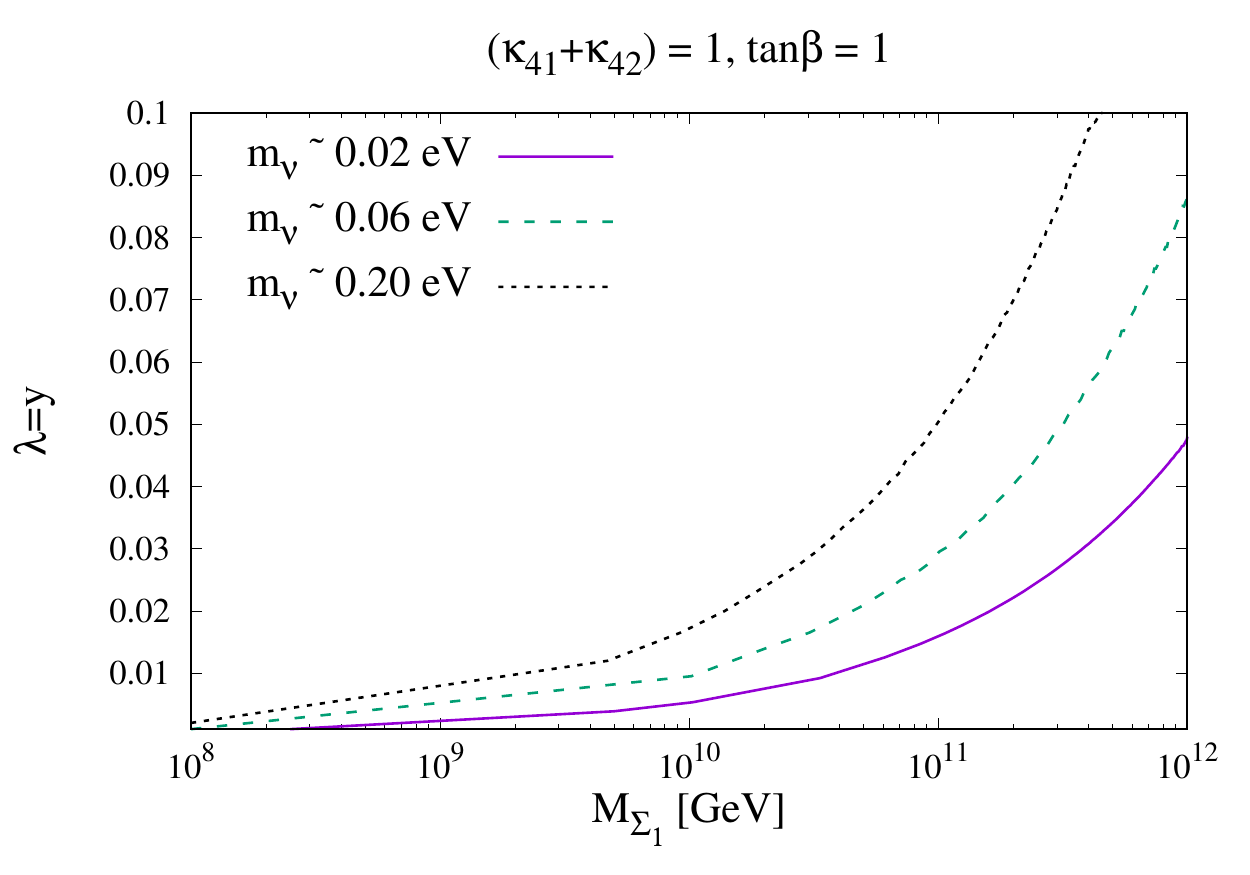}}
\subfigure []{
\includegraphics[height=6 cm, width=7 cm,angle=0]{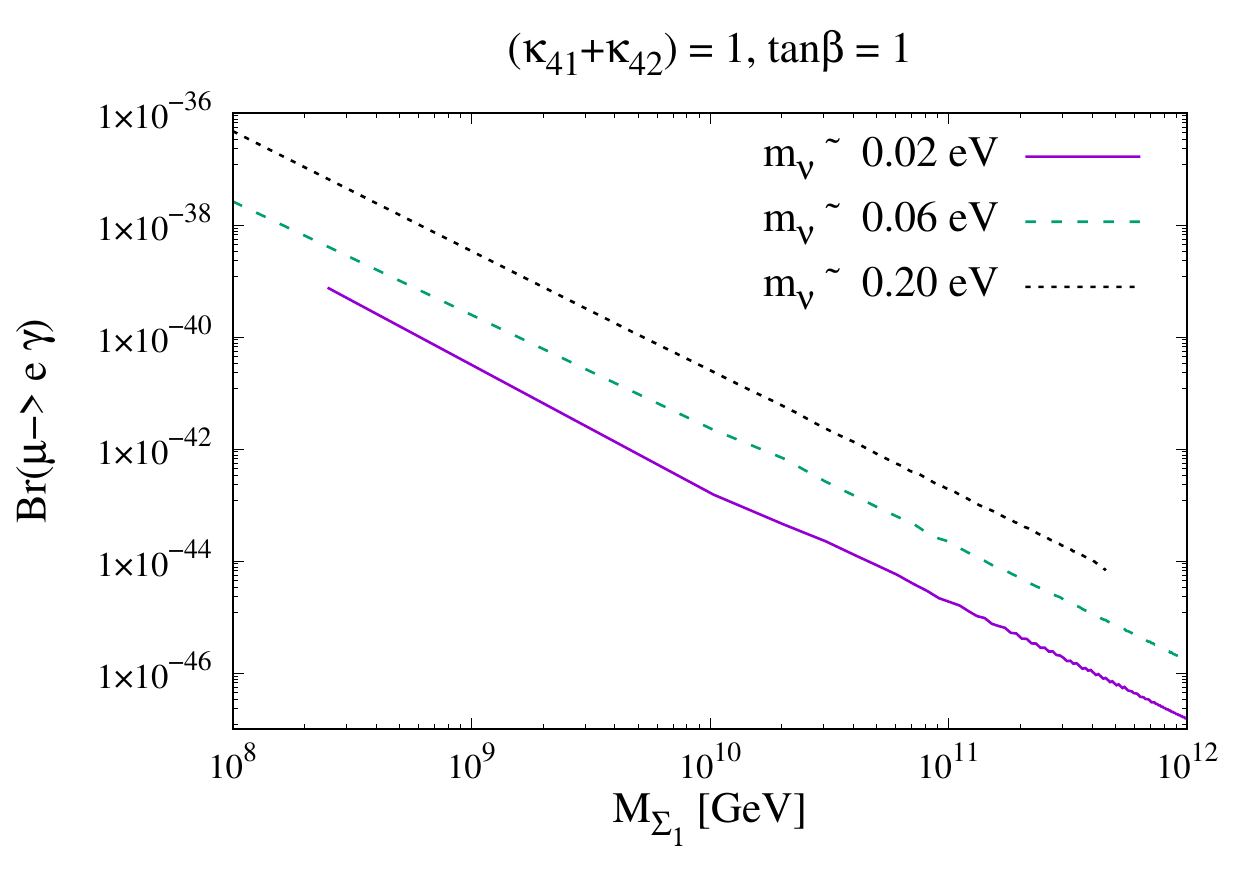}}
 \caption{Left panel: Variation of $M_{\Sigma_1}$ with Yukawa coupling $\lambda$ consistent with active  neutrino mass constraints assuming $\lambda=y$ for $\tan\beta=1$ and $|\kappa_{41,42}|=0.5$.  Right panel: Decay branching ratio ${\mathcal B}(\mu \rightarrow e \gamma)$ versus $M_{\Sigma_1}$ for the  allowed range of $\lambda=y$ coupling in agreement with limit from neutrino mass.}
  \label{meg}
\end{figure}

In Fig.~\ref{meg}(a), we present the variation of vector fermion mass $M_{\Sigma_1}$ with Yukawa couplings imposing the condition $\lambda=y$ for neutrino Yukawa couplings. The mass scale of light active neutrino mass $(m_\nu)$ is derived using Eq.~(\ref{nmass4}) for $\tan\beta=1$ and $|\kappa_{41,42}|=0.5$ with assuming a hierarchical structure of vector fermion mass $M_{\Sigma_1}: M_{\Sigma_2}: M_{\Sigma_3}=1:3:30$.  We further consider mass of different inert scalar $m_{S}\sim1.5$ TeV, following results from Fig.~\ref{fig5} and Fig.~\ref{fig8}, consistent with relic density, direct detection, $T$ parameter and collider constraints on dark matter. 
From Fig.~\ref{meg}(a), we notice that for Yukawa coupling ranging from $10^{-3}-10^{-1}$, mass of vector fermion varies within the range $10^{8}-10^{12}$ GeV. Using the allowed range of $M_{\Sigma_1}$ and Yukawa coupling values, we plot the corresponding decay branching ratio ${\mathcal B}(\mu \rightarrow e \gamma)$ against $M_{\Sigma_1}$, as shown in Fig.~\ref{meg}(b). It is clearly observed that within the specific range of $\lambda (y)-M_{\Sigma_1}$, $\mu \rightarrow e \gamma$ decay branching ratio is negligible compared to the experimental limit ${\mathcal B}(\mu \rightarrow e \gamma) < 4.2 \times 10^{-13}$~\cite{TheMEG:2016wtm}. Therefore, lepton flavor violation do not impose any stringent limit in the present framework.

\subsection*{Leptogenesis in sGM model}

Apart from the generation of one-loop neutrino mass and providing a feasible candidate for dark matter, the scotogenic GM model can also generate lepton asymmetry which can explain baryon asymmetry in the Universe (BAU). In the present model, heavy vector-like fermions can decay into leptons and triplets. A net asymmetry in lepton number can be generated from the CP violating decay of heavy vector-like fermions. 

The net amount of asymmetry generated from CP violating decay of VLF's coming from its decay into leptons with two different scalar triplets are expressed as
\be
\epsilon_{1\Delta}  =  \frac{\sum_{\alpha}[\Gamma(\Sigma \rightarrow l^c_L+\Delta)-
\Gamma(\Sigma^c \rightarrow {l}_{L}+\Delta^{*})]}{\Gamma_1}\,\, ,
\label{asyD}
\ee
and
\be
\epsilon_{1T}  =  \frac{\sum_{\alpha}[\Gamma(\Sigma^c \rightarrow l^c_L+T^*)-
\Gamma(\Sigma \rightarrow {l}_{L}+T)]}{\Gamma_1}\,\, .
\label{asyT}
\ee
In the above Eq.~(\ref{asyD})-(\ref{asyT}), total decay width of vector fermion is given as 
\be
{\Gamma_1}=\frac{3}{32\pi}\left((y^{\dagger}y)_{11}+(\lambda^{\dagger}\lambda)_{11}\right) M_{\Sigma_1}\, .
\label{decay}
\ee
The asymmetry $\epsilon_{1\Delta}$ can be redefined as
\be
\epsilon_{1\Delta}=
-\frac{3}{8\pi}\frac{1}{(y^{\dagger}y)_{11}+(\lambda^{\dagger}\lambda)_{11}}\sum_{j}{\rm Im}
[(y^{\dagger}y)_{1j}(\lambda^{T}\lambda^*)_{j1}]\frac{M_{\Sigma_1}}{M_{\Sigma_j}}\,\, .
\label{asyd2}
\ee
Similarly, the asymmetry  $\epsilon_{1T}$ becomes  
\be
\epsilon_{1T}=
-\frac{3}{8\pi}\frac{1}{(y^{\dagger}y)_{11}+(\lambda^{\dagger}\lambda)_{11}}\sum_{j}{\rm Im}
[(\lambda^{\dagger}\lambda)_{1j}(y^{T}y^*)_{j1}]\frac{M_{\Sigma_1}}{M_{\Sigma_j}}\,\, .
\label{asyT2}
\ee 

As mentioned before, in the following study we consider the mass matrix of vector-like fermions to be diagonal and assume hierarchical structure of vector-like fermion masses such that $M_{\Sigma_1}<M_{\Sigma_2}<M_{\Sigma_3}$ and $M_{\Sigma_1}:M_{\Sigma_2}:M_{\Sigma_3}=1:3:30$. Therefore, Boltzmann equations for leptogenesis consist of three equations (which is governed by decay of lightest vector fermion) 
\begin{eqnarray}
\label{eq:BE}
\frac{d Y_{\Sigma_1}}{dz}&=& -z\frac{\Gamma_1}{{\rm H_1}} 
\frac{K_1(z)}{K_2(z)}\left(Y_{\Sigma_1}-Y_{\Sigma_1}^{\rm eq}\right) \, \,,
\\ 
\label{eq:BE1}
\frac{d Y_{L}^{\Delta}}{dz} &=&  - \frac{\Gamma_1}{{\rm H_1}} 
\left (\epsilon_{1\Delta}  z \frac{K_1(z)}{K_2(z)}(Y_{\Sigma_1}^{eq}- Y_{\Sigma_1})  +  Br_L^{\Delta}  
\frac{z^3 K_1(z)}{4} Y_{L}^{\Delta}  \right)\,, \,
\end{eqnarray}
and 
 \begin{eqnarray}
  \label{eq:BE2}
\frac{d Y_{L}^{T}}{dz} &=&  - \frac{\Gamma_1}{{\rm H_1}} 
\left (\epsilon_{1T}  z \frac{K_1(z)}{K_2(z)}(Y_{\Sigma_1}^{eq}- Y_{\Sigma_1})  +  Br_L^{T}  
\frac{z^3 K_1(z)}{4} Y_{L}^{T}  \right)\,, \,
 \end{eqnarray}
where  Eq.~(\ref{eq:BE}) denotes Boltzmann equation for decay of vector-like fermion $\Sigma_1$, other two Eqs.~(\ref{eq:BE1})-(\ref{eq:BE2}) are Boltzmann equations for generation of lepton asymmetry. In Eqs.~(\ref{eq:BE})-(\ref{eq:BE2}), ${\rm H_1}$ denotes the Hubble parameter at $T=M_{\Sigma_1}$ and $Y_{x}={n_x }/{s}$;  ($x=\Sigma_1, L$, $n$: number density and $s:$ co-moving entropy density).  The expression for ${\rm H_1}$ and  equilibrium yield of $\Sigma_1$ are 
\be
{\rm H_1}=\sqrt{\frac{8\pi G \rho_{rad}}{3}}=1.66 \sqrt{g_*} \frac{M_{\Sigma_1}^2}{M_P}\,, 
\hskip 5mm
Y_{\Sigma_1}^{\rm eq}=\frac{45g}{4\pi^4}\frac{z^2K_2(z)}{g_{*s}}\,,
\label{Hub}
\ee
where $g_*=119.75$ (including new scalar fields) is the relativistic degrees of freedom and $g_{*s}$ is entropy degrees of freedom respectively. Factors $K_{1,2}$ in Eqs.~(\ref{eq:BE1})-(\ref{eq:BE2}) are modified Bessel functions and $Br_{L}^x\,(x=\Delta\,,T)$ represent the decay branching fraction of $\Sigma_1$ into $\Delta$ and $T$ mode such that
\be
Br_{L}^{\Delta}=\frac{(y^{\dagger}y)_{11}}{(y^{\dagger}y)_{11}+(\lambda^{\dagger}\lambda)_{11}}\, 
,\hskip 5mm
Br_{L}^{T}=\frac{(\lambda^{\dagger}\lambda)_{11}}{(y^{\dagger}y)_{11}+(\lambda^{\dagger}\lambda)_{11}}\, .
\label{branch}
\ee
It is to be noted that Boltzmann Eqs.~(\ref{eq:BE1})-(\ref{eq:BE2}) are based on the assumption that the transfer of asymmetry between $Y_{L}^T$ and $Y_{L}^{\Delta}$ is negligible and their evolution are independent of each other. This is possible when the condition for the narrow width approximation is satisfied which is given as~\cite{Falkowski:2011xh}
\be
\frac{\Gamma_1}{{\rm H_1}} \frac{\Gamma_1}{M_{\Sigma_1}} < 0.1 \, . 
\label{nwidth}
\ee
The final lepton asymmetry obtained by solving Boltzmann equations for leptogenesis is given as
\be
Y_{L}(z\rightarrow \infty)= 2\left(Y_{L}^{\Delta}(z\rightarrow \infty) + Y_{L}^{T}(z\rightarrow 
\infty)\right)\,\, .
\label{lepasy}
\ee
In the above equation the additional factor of two is due to the fact that asymmetry is generated from both neutral and charged fermions of the vector fermion doublet. Eq.~(\ref{lepasy}) reveals that, there can be a cancellation in net asymmetry $Y_{L}(z\rightarrow \infty)$. Similar to the case of standard leptogenesis with right handed neutrinos, here the final lepton asymmetry is also partially transferred into baryon asymmetry via sphaleron transition~\cite{Davidson:2008bu}
\be
Y_{B}(z\rightarrow \infty)= \frac{10}{31} Y_{L}(z\rightarrow \infty)\,\, .
\ee
Baryon asymmetry of the Universe as measured by PLANCK is $Y_B=(8-9.5)\times10^{-11}$, hence the required  lepton asymmetry is $Y_L\sim(2.5-2.9)\times10^{-10}$~\cite{PhysRevD.98.030001}.
\begin{table}[htb]
 \begin{center}
 \begin{tabular}{|c| c|c| c| c|c|c|c|} 
 \hline
 \hline
~ Set~ & $\lambda=y$ & $M_{\Sigma_1} (\rm GeV)$ & ${\mathcal O}(m_\nu)~({\rm eV})$ &
$\epsilon_{1T,1\Delta}$ & $Br_L^{T}=Br_L^{\Delta}$ & ~$\frac{\Gamma_1}{{\rm H_1}} \frac{\Gamma_1}{M_{\Sigma_1}}$~ &~$Y_L(z\rightarrow \infty)$~\\ [0.5ex] 
 \hline
 I& $4.0\times10^{-2}$& $6.75\times10^{11}$ & $ 0.02$ &$9.53\times10^{-5}$   & 0.5 & $8.1\times10^{-2}$&$4.13\times10^{-9}$  \\ 
 \hline
 II& $2.1\times10^{-2}$  & $5.0\times10^{10}$ & $ 0.06$ &$2.63\times10^{-5}$   & 0.5 & $8.3\times10^{-2}$& $2.61\times10^{-10}$ \\ 
 \hline
  III& $1.1\times10^{-2}$  & $4.0\times10^{9}$ & $ 0.20$ &$7.19\times10^{-6}$   & 0.5 & $7.8\times10^{-2}$&$1.92\times10^{-11  } $ \\ 
 \hline
 \hline
 \end{tabular}
\end{center}
\caption{ Lepton asymmetry $Y_L$ calculated using given set of vector fermion $M_{\Sigma_1}$ mass and 
Yukawa coupling for $(\kappa_{41}+\kappa_{42})=1$, $\tan\beta=1$ with different sets of Yukawa couplings 
$\{\lambda, y\}$  for  neutrino mass scale $m_{\nu}\sim {\mathcal O}(10^{-2}-10^{-1})$ eV.}
\label{tab:4}
\end{table} 

\begin{figure}[!ht]
\centering
\subfigure[]{
\includegraphics[height=6 cm, width=7 cm,angle=0]{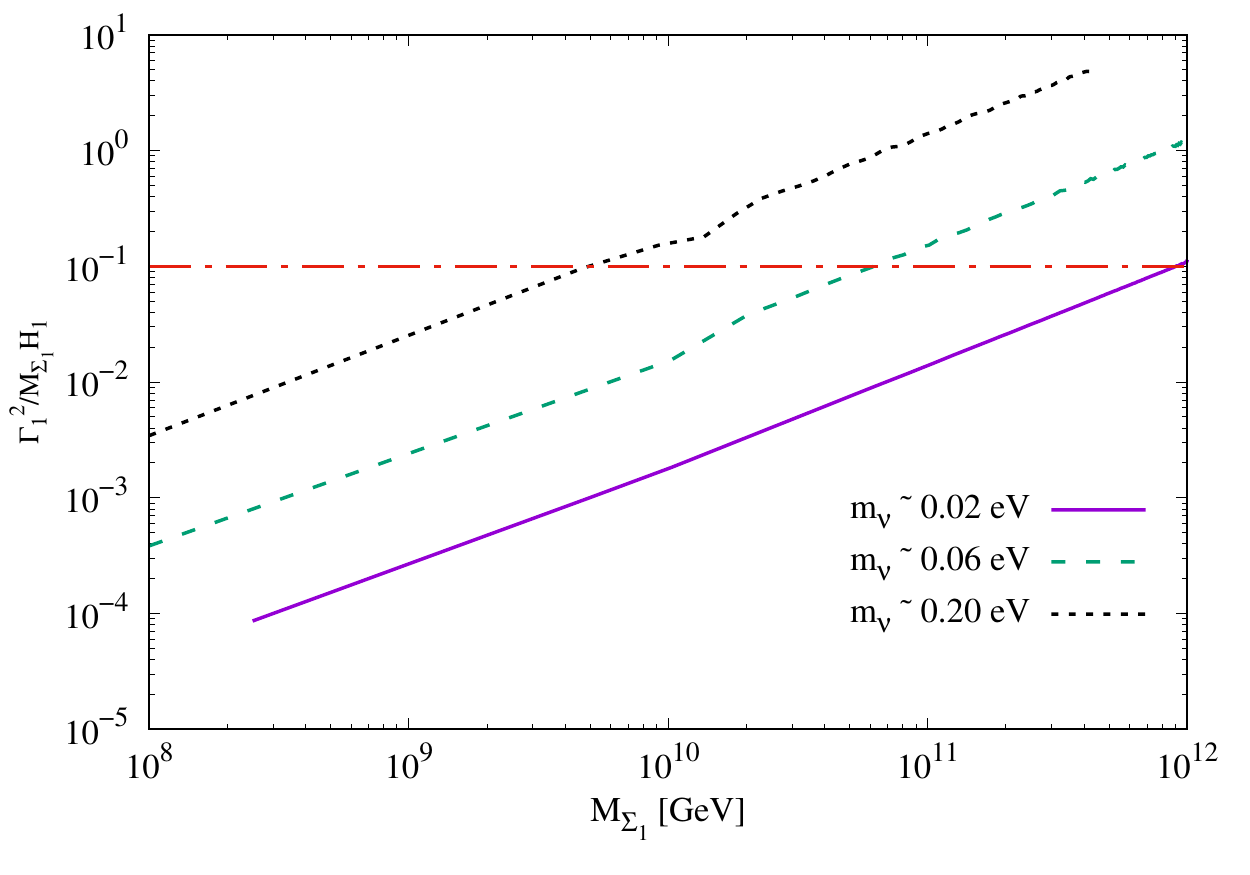}}
\subfigure []{
\includegraphics[height=6 cm, width=7 cm,angle=0]{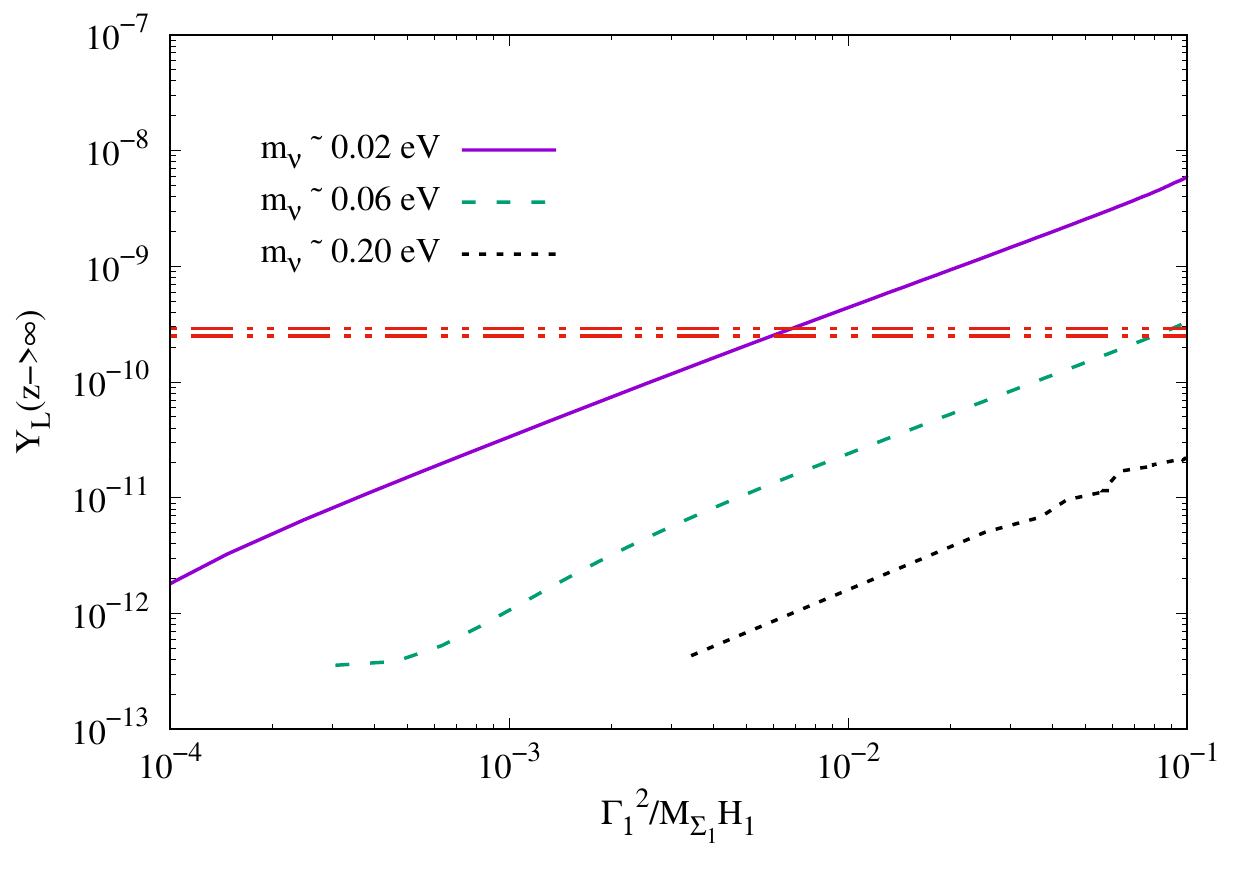}}
\caption{Variation of $M_{\Sigma_1}$ against ${\Gamma_1}^{2}/M_{\Sigma_1}{\rm H_1}$
(left panel) and corresponding lepton asymmetry $Y_L$ for 
${\Gamma_1}^{2}/M_{\Sigma_1}{\rm H_1}<0.1$ (right panel).}
\label{fig9}
\end{figure}

We now investigate the possibility whether the present model can generate the required BAU. In order to do so, we first consider Yukawa couplings and  mass of vector-like fermions that will provide sub-eV light neutrino mass. For this purpose, we use the allowed $M_{\Sigma_1}$ and Yukawa coupling $\lambda=y$ parameter space derived in Fig.~\ref{meg}(a) for neutrino mass scale $m_{\nu}=0.02-0.20$ eV.  Moreover, the Boltzmann equations for lepton asymmetry used are valid only when the condition for narrow width approximation in Eq.~(\ref{nwidth}) is respected. It is found that the washout and transfer of asymmetry becomes significant for ${\Gamma_1}^{2}/M_{\Sigma_1}{\rm H_1}\geq 0.1 $~\cite{Falkowski:2011xh}. To circumvent this issue, we further restrict the allowed $M_{\Sigma_1}$ vs $\lambda = y $ parameter space in Fig.~\ref{meg}(a) with the condition ${\Gamma_1}^{2}/M_{\Sigma_1}{\rm H_1}< 0.1$. In Fig.~\ref{fig9}(a), we present the variation of vector fermion mass $M_{\Sigma_1}$ against the narrow width approximation parameter using the neutrino mass constraints from Fig.~\ref{meg}(a). The upper limit for narrow width approximation ${\Gamma_1}^{2}/M_{\Sigma_1}{\rm H_1}< 0.1$ is represented by the red horizontal line shown in Fig.~\ref{fig9}(a). With the given range of neutrino mass scale $m_\nu=0.02-0.20$ eV, Fig.~\ref{fig9}(a) provides an upper limit on vector  fermion mass for which narrow width approximation and the Boltzmann equations (Eqs.~(\ref{eq:BE1})-(\ref{eq:BE2})) for lepton asymmetry remains valid. 

Using the allowed range of $M_{\Sigma_1}$ values obtained from Fig.~\ref{fig9}(a) and Yukawa coupling from Fig.~\ref{meg}(a), we solve  Boltzmann equations Eqs.~(\ref{eq:BE})-(\ref{eq:BE2}) and evaluate the  amount of lepton asymmetry in the present formalism. As to obtain the limits in  Fig.~\ref{meg}(a) and  Fig.~\ref{fig9}(a), taking the condition $\lambda=y$ we get $\epsilon_{1\Delta}=\epsilon_{1T}$ and $Br_L^T=Br_L^{\Delta}=1/2$. Therefore, for $\lambda=y$,  lepton asymmetry generated from Eqs.~(\ref{eq:BE1})-(\ref{eq:BE2}) will be equal, i.e.; $Y_L^{\Delta}(z\rightarrow \infty)=Y_L^T(z\rightarrow \infty)$. In Fig.~\ref{fig9}(b), we plot the total lepton asymmetry $Y_L$ versus  ${\Gamma_1}^{2}/M_{\Sigma_1}{\rm H_1}$ using the available $M_{\Sigma_1}$ parameter space obtained for aforementioned $m_\nu$ values. Horizontal red lines in Fig.~\ref{fig9}(b) exhibit the $Y_L$ value required to generate observed baryon asymmetry in the Universe. For demonstrative purpose, in Table~\ref{tab:4}, we tabulate lepton asymmetry  $Y_L$ for few sets of benchmark parameters obtained from the solutions of Boltzmann equations for leptogenesis. We observe that for benchmark set-I, yield of lepton asymmetry $Y_L$ is above the required amount of matter-antimatter asymmetry. On the other hand, for benchmark set-II with $m_\nu \sim 0.06$ eV, net lepton asymmetry $Y_L$ successfully generates observed baryon abundance. However, set-III result in Table~\ref{tab:4} suggests that yield $Y_L$  fails to explain baryon asymmetry in the Universe. 

It is worth mentioning that solution to  Boltzmann equations in this work is based on the assumption that the Universe is radiation dominated at the era of leptogenesis. However, this situation can alter if one considers a modified thermal history of Universe~\cite{DEramo:2017gpl}. Recent studies~\cite{Dutta:2018zkg,Chen:2019etb,Mahanta:2019sfo,Konar:2020vuu} reveal that non-standard cosmological effects (such as a new scalar field active near the temperature $T\sim M_{\Sigma_1}$) on leptogenesis can reduce the effects of washout significantly and enhance the yield of $Y_L$ by one or two order depending on the new parameters. We expect similar changes in lepton asymmetry  $Y_L$  if non-standard cosmological effects are taken into account. In such scenarios, benchmark set-III of Table~\ref{tab:4} can also explain matter-antimatter asymmetry in the Universe. Therefore, non-standard leptogenesis scenarios can relax constraints  and  broaden the available model parameter space considered in the model. However at lower temperature $T\sim10^{2}-10^{3}$ TeV, the non-standard effect is completely washed out and Universe becomes radiation dominated. Therefore, thermal freeze-out of dark  matter remains unaffected and hence dark matter phenomenology discussed in the work remains unharmed.

\subsection*{Conclusions}

In this work we present a study of common origin of neutrino mass, dark matter and leptogenesis by extending Georgi-Machacek model with two Higgs doublet and new vector fermions. The composition of Georgi-Machacek model includes two additional $SU(2)$ triplet scalar with different hypercharge along with the SM Higgs doublet. Although the GM model preserves custodial symmetry and can generate neutrino mass in tree level, the model cannot account for a stable dark matter candidate. 
We consider a scalar potential invariant under $SU(2)_L\times U(1)_Y\times \mathcal {Z}_4 \times \mathcal {Z}_2$.
After spontaneous breaking of symmetry, $\mathcal {Z}_2$ symmetry of the potential remains conserved by triplet fields and vector fermions. The $\mathcal {Z}_4$ symmetry in the potential provides necessary quartic interactions that generate neutrino mass at one-loop upon mixing of neutral and charged scalar of $Y=0$ and $Y=1$ triplet.
Thus a residual $\mathcal {Z}_2$ symmetry ensures the stability of neutral scalar originating from the mixing of scalar triplets which serves the purpose of dark matter resulting a scotogenic Georgi-Machacek (sGM )model.
 Moreover, the sGM model also exhibits feature of leptogenesis, capable of generating matter-antimatter asymmetry in the Universe from decaying vector fermions.

We put constraints on the model parameter space from various theoretical and experimental observations, such as vacuum stability, dark matter relic abundance,  dark matter direct detection cross-section. Detailed study of dark matter phenomenology reveals that  the model predicts a TeV scale dark matter candidate. In the present model, matter-antimatter asymmetry originates from CP violating decay of vector-like fermions. Using the limits from light neutrino mass, we found that massive vector fermions around $10^{10-11}$ GeV can produce the experimentally observed baryon abundance. We have also observed that lepton flavor violation do not impose any significant constraint in the model parameter space. Therefore, a simple extension of Georgi-Machacek model with two Higgs doublet and heavy vector-like fermions accompanied by $\mathcal {Z}_4$ can provide answers to the puzzles of dark matter, matter-antimatter asymmetry in a single framework with radiative generation of neutrino mass. 

~\\
\noindent\textbf{Acknowledgements}

This work is supported in part by the National Science Foundation of China (11775093, 11422545, 11947235).

\appendix
\section{Appendix: SM Higgs coupling with charged scalars}
The couplings of SM Higgs $h$ with different charged scalars are listed as follows:
\begin{align*}
\lambda_{hD^{++}D^{--}} &=  s_{\alpha} c_{\beta} \kappa_{11} - c_{\alpha} s_{\beta} \kappa_{12}\,, \\ 	
\lambda_{hS_1^+S_1^-} &=
	\frac{1}{2}\big(2 s_{\delta}^2  s_{\alpha} c_{\beta}\kappa_{11}-2 s_{\delta}^2  c_{\alpha} s_{\beta}\kappa_{12}+ s_{\delta}^2  s_{\alpha} c_{\beta}\kappa_{21} \\ 
  & - s_{\delta}^2  c_{\alpha} s_{\beta}\kappa_{22}+2 c_{\delta}^2  s_{\alpha} c_{\beta}\kappa_{31}-2 c_{\delta}^2  c_{\alpha}s_{\beta}\kappa_{32} -\sqrt{2}  c_{\delta} s_{\delta}c_{\alpha} c_{\beta}\kappa_{41} \\
  & + \sqrt{2}c_{\delta} s_{\delta} s_{\alpha} s_{\beta}\kappa_{41}+ \sqrt{2} c_{\delta} s_{\delta}s_{\alpha} s_{\beta}\kappa_{42}- \sqrt{2}c_{\delta} s_{\delta} c_{\alpha} c_{\beta}\kappa_{42}\big)\,,\\
\lambda_{hS_2^+S_2^-} &=
	\frac{1}{2}\big(2 c_{\delta}^2  s_{\alpha} c_{\beta}\kappa_{11}-2 c_{\delta}^2  c_{\alpha} s_{\beta}\kappa_{12}+ c_{\delta}^2  s_{\alpha} c_{\beta}\kappa_{21}
  - c_{\delta}^2  c_{\alpha} s_{\beta}\kappa_{22} \\ & +2 s_{\delta}^2  s_{\alpha}c_{\beta}\kappa_{31} -2 s_{\delta}^2  c_{\alpha} s_{\beta}\kappa_{32}+ \sqrt{2}c_{\delta} s_{\delta} c_{\alpha} c_{\beta}\kappa_{41} \\
  & - \sqrt{2}c_{\delta} s_{\delta} s_{\alpha} s_{\beta}\kappa_{41}- \sqrt{2}c_{\delta} s_{\delta} s_{\alpha} s_{\beta}\kappa_{42}+ \sqrt{2}c_{\delta} s_{\delta} c_{\alpha} c_{\beta}\kappa_{42}\big)\,,\\ 
\lambda_{hH^+H^-}  &=
	 s_{\alpha}s_{\beta}^2  c_{\beta} \lambda_1- c_{\alpha}c_{\beta}^2   s_{\beta}\lambda_2- c_{\alpha}s_{\beta}^3\lambda_3   + c_{\beta}^3   s_{\alpha}\lambda_3  \\
   & - c_{\beta}  s_{\alpha}s_{\beta}^2\lambda_4+ c_{\alpha}c_{\beta}^2  s_{\beta}\lambda_4+ c_{\alpha}c_{\beta}^2 s_{\beta} \lambda_5 - c_{\beta}  s_{\alpha}s_{\beta}^2\lambda_5\,.
  \end{align*}
   
\bibliography{reference2}

\begin{thebibliography}{10}
\expandafter\ifx\csname url\endcsname\relax
  \def\url#1{\texttt{#1}}\fi
\expandafter\ifx\csname urlprefix\endcsname\relax\def\urlprefix{URL }\fi
\expandafter\ifx\csname href\endcsname\relax
  \def\href#1#2{#2} \def\path#1{#1}\fi

\bibitem{Aghanim:2018eyx}
N.~Aghanim, et~al., {Planck 2018 results. VI. Cosmological parameters}\href
  {http://arxiv.org/abs/1807.06209} {\path{arXiv:1807.06209}}.

\bibitem{PhysRevD.98.030001}
M.~Tanabashi, e.~a. Hagiwara, Review of particle physics, Phys. Rev. D 98
  (2018) 030001.
\newblock \href {http://dx.doi.org/10.1103/PhysRevD.98.030001}
  {\path{doi:10.1103/PhysRevD.98.030001}}.

\bibitem{Minkowski:1977sc}
P.~Minkowski, {$\mu \to e\gamma$ at a Rate of One Out of $10^{9}$ Muon
  Decays?}, Phys. Lett. B 67 (1977) 421--428.
\newblock \href {http://dx.doi.org/10.1016/0370-2693(77)90435-X}
  {\path{doi:10.1016/0370-2693(77)90435-X}}.

\bibitem{GellMann:1980vs}
M.~Gell-Mann, P.~Ramond, R.~Slansky, {Complex Spinors and Unified Theories},
  Conf. Proc. C 790927 (1979) 315--321.
\newblock \href {http://arxiv.org/abs/1306.4669} {\path{arXiv:1306.4669}}.

\bibitem{Mohapatra:1979ia}
R.~N. Mohapatra, G.~Senjanovic, {Neutrino Mass and Spontaneous Parity
  Nonconservation}, Phys. Rev. Lett. 44 (1980) 912.
\newblock \href {http://dx.doi.org/10.1103/PhysRevLett.44.912}
  {\path{doi:10.1103/PhysRevLett.44.912}}.

\bibitem{Schechter:1980gr}
J.~Schechter, J.~Valle, {Neutrino Masses in SU(2) x U(1) Theories}, Phys. Rev.
  D 22 (1980) 2227.
\newblock \href {http://dx.doi.org/10.1103/PhysRevD.22.2227}
  {\path{doi:10.1103/PhysRevD.22.2227}}.

\bibitem{Mohapatra:1980yp}
R.~N. Mohapatra, G.~Senjanovic, {Neutrino Masses and Mixings in Gauge Models
  with Spontaneous Parity Violation}, Phys. Rev. D 23 (1981) 165.
\newblock \href {http://dx.doi.org/10.1103/PhysRevD.23.165}
  {\path{doi:10.1103/PhysRevD.23.165}}.

\bibitem{Wetterich:1981bx}
C.~Wetterich, {Neutrino Masses and the Scale of B-L Violation}, Nucl. Phys. B
  187 (1981) 343--375.
\newblock \href {http://dx.doi.org/10.1016/0550-3213(81)90279-0}
  {\path{doi:10.1016/0550-3213(81)90279-0}}.

\bibitem{Schechter:1981cv}
J.~Schechter, J.~Valle, {Neutrino Decay and Spontaneous Violation of Lepton
  Number}, Phys. Rev. D 25 (1982) 774.
\newblock \href {http://dx.doi.org/10.1103/PhysRevD.25.774}
  {\path{doi:10.1103/PhysRevD.25.774}}.

\bibitem{Hambye:2003ka}
T.~Hambye, G.~Senjanovic, {Consequences of triplet seesaw for leptogenesis},
  Phys. Lett. B 582 (2004) 73--81.
\newblock \href {http://arxiv.org/abs/hep-ph/0307237}
  {\path{arXiv:hep-ph/0307237}}, \href
  {http://dx.doi.org/10.1016/j.physletb.2003.11.061}
  {\path{doi:10.1016/j.physletb.2003.11.061}}.

\bibitem{Antusch:2007km}
S.~Antusch, {Flavour-dependent type II leptogenesis}, Phys. Rev. D 76 (2007)
  023512.
\newblock \href {http://arxiv.org/abs/0704.1591} {\path{arXiv:0704.1591}},
  \href {http://dx.doi.org/10.1103/PhysRevD.76.023512}
  {\path{doi:10.1103/PhysRevD.76.023512}}.

\bibitem{Foot:1988aq}
R.~Foot, H.~Lew, X.~He, G.~C. Joshi, {Seesaw Neutrino Masses Induced by a
  Triplet of Leptons}, Z. Phys. C 44 (1989) 441.
\newblock \href {http://dx.doi.org/10.1007/BF01415558}
  {\path{doi:10.1007/BF01415558}}.

\bibitem{Chen:2009vx}
S.-L. Chen, X.-G. He, {Leptogenesis and LHC Physics with Type III See-Saw},
  Int. J. Mod. Phys. Conf. Ser. 01 (2011) 18--27.
\newblock \href {http://arxiv.org/abs/0901.1264} {\path{arXiv:0901.1264}},
  \href {http://dx.doi.org/10.1142/S2010194511000067}
  {\path{doi:10.1142/S2010194511000067}}.

\bibitem{Fukugita:1986hr}
M.~Fukugita, T.~Yanagida, {Baryogenesis Without Grand Unification}, Phys. Lett.
  B 174 (1986) 45--47.
\newblock \href {http://dx.doi.org/10.1016/0370-2693(86)91126-3}
  {\path{doi:10.1016/0370-2693(86)91126-3}}.

\bibitem{Buchmuller:2004nz}
W.~Buchmuller, P.~Di~Bari, M.~Plumacher, {Leptogenesis for pedestrians}, Annals
  Phys. 315 (2005) 305--351.
\newblock \href {http://arxiv.org/abs/hep-ph/0401240}
  {\path{arXiv:hep-ph/0401240}}, \href
  {http://dx.doi.org/10.1016/j.aop.2004.02.003}
  {\path{doi:10.1016/j.aop.2004.02.003}}.

\bibitem{Buchmuller:2005eh}
W.~Buchmuller, R.~Peccei, T.~Yanagida, {Leptogenesis as the origin of matter},
  Ann. Rev. Nucl. Part. Sci. 55 (2005) 311--355.
\newblock \href {http://arxiv.org/abs/hep-ph/0502169}
  {\path{arXiv:hep-ph/0502169}}, \href
  {http://dx.doi.org/10.1146/annurev.nucl.55.090704.151558}
  {\path{doi:10.1146/annurev.nucl.55.090704.151558}}.

\bibitem{Davidson:2008bu}
S.~Davidson, E.~Nardi, Y.~Nir, {Leptogenesis}, Phys. Rept. 466 (2008) 105--177.
\newblock \href {http://arxiv.org/abs/0802.2962} {\path{arXiv:0802.2962}},
  \href {http://dx.doi.org/10.1016/j.physrep.2008.06.002}
  {\path{doi:10.1016/j.physrep.2008.06.002}}.

\bibitem{An:2009vq}
H.~An, S.-L. Chen, R.~N. Mohapatra, Y.~Zhang, {Leptogenesis as a Common Origin
  for Matter and Dark Matter}, JHEP 03 (2010) 124.
\newblock \href {http://arxiv.org/abs/0911.4463} {\path{arXiv:0911.4463}},
  \href {http://dx.doi.org/10.1007/JHEP03(2010)124}
  {\path{doi:10.1007/JHEP03(2010)124}}.

\bibitem{Ma:2006km}
E.~Ma, {Verifiable radiative seesaw mechanism of neutrino mass and dark
  matter}, Phys. Rev. D 73 (2006) 077301.
\newblock \href {http://arxiv.org/abs/hep-ph/0601225}
  {\path{arXiv:hep-ph/0601225}}, \href
  {http://dx.doi.org/10.1103/PhysRevD.73.077301}
  {\path{doi:10.1103/PhysRevD.73.077301}}.

\bibitem{Kashiwase:2012xd}
S.~Kashiwase, D.~Suematsu, {Baryon number asymmetry and dark matter in the
  neutrino mass model with an inert doublet}, Phys. Rev. D 86 (2012) 053001.
\newblock \href {http://arxiv.org/abs/1207.2594} {\path{arXiv:1207.2594}},
  \href {http://dx.doi.org/10.1103/PhysRevD.86.053001}
  {\path{doi:10.1103/PhysRevD.86.053001}}.

\bibitem{Kashiwase:2013uy}
S.~Kashiwase, D.~Suematsu, {Leptogenesis and dark matter detection in a TeV
  scale neutrino mass model with inverted mass hierarchy}, Eur. Phys. J. C 73
  (2013) 2484.
\newblock \href {http://arxiv.org/abs/1301.2087} {\path{arXiv:1301.2087}},
  \href {http://dx.doi.org/10.1140/epjc/s10052-013-2484-9}
  {\path{doi:10.1140/epjc/s10052-013-2484-9}}.

\bibitem{Hugle:2018qbw}
T.~Hugle, M.~Platscher, K.~Schmitz, {Low-Scale Leptogenesis in the Scotogenic
  Neutrino Mass Model}, Phys. Rev. D 98~(2) (2018) 023020.
\newblock \href {http://arxiv.org/abs/1804.09660} {\path{arXiv:1804.09660}},
  \href {http://dx.doi.org/10.1103/PhysRevD.98.023020}
  {\path{doi:10.1103/PhysRevD.98.023020}}.

\bibitem{Lu:2016dbc}
W.-B. Lu, P.-H. Gu, {Mixed Inert Scalar Triplet Dark Matter, Radiative Neutrino
  Masses and Leptogenesis}, Nucl. Phys. B 924 (2017) 279--311.
\newblock \href {http://arxiv.org/abs/1611.02106} {\path{arXiv:1611.02106}},
  \href {http://dx.doi.org/10.1016/j.nuclphysb.2017.09.005}
  {\path{doi:10.1016/j.nuclphysb.2017.09.005}}.

\bibitem{Cao:2017xgk}
Q.-H. Cao, S.-L. Chen, E.~Ma, B.~Yan, D.-M. Zhang, {New Class of Two-Loop
  Neutrino Mass Models with Distinguishable Phenomenology}, Phys. Lett. B 779
  (2018) 430--435.
\newblock \href {http://arxiv.org/abs/1707.05896} {\path{arXiv:1707.05896}},
  \href {http://dx.doi.org/10.1016/j.physletb.2018.02.038}
  {\path{doi:10.1016/j.physletb.2018.02.038}}.

\bibitem{Zhou:2017lrt}
H.~Zhou, P.-H. Gu, {From high-scale leptogenesis to low-scale one-loop neutrino
  mass generation}, Nucl. Phys. B 927 (2018) 184--195.
\newblock \href {http://arxiv.org/abs/1708.04207} {\path{arXiv:1708.04207}},
  \href {http://dx.doi.org/10.1016/j.nuclphysb.2017.12.016}
  {\path{doi:10.1016/j.nuclphysb.2017.12.016}}.

\bibitem{Gu:2018kmv}
P.-H. Gu, H.-J. He, {TeV Scale Neutrino Mass Generation, Minimal Inelastic Dark
  Matter, and High Scale Leptogenesis}, Phys. Rev. D 99~(1) (2019) 015025.
\newblock \href {http://arxiv.org/abs/1808.09377} {\path{arXiv:1808.09377}},
  \href {http://dx.doi.org/10.1103/PhysRevD.99.015025}
  {\path{doi:10.1103/PhysRevD.99.015025}}.

\bibitem{DuttaBanik:2020vfr}
A.~Dutta~Banik, R.~Roshan, A.~Sil, {Neutrino mass and asymmetric dark matter:
  study with inert Higgs doublet and high scale validity}\href
  {http://arxiv.org/abs/2011.04371} {\path{arXiv:2011.04371}}.

\bibitem{Lineros:2020eit}
R.~A. Lineros, M.~Pierre, {Dark Matter candidates in a Type-II radiative
  neutrino mass model}\href {http://arxiv.org/abs/2011.08195}
  {\path{arXiv:2011.08195}}.

\bibitem{HABER1979493}
H.~Haber, G.~Kane, T.~Sterling, The fermion mass scale and possible effects of
  higgs bosons on experimental observables, Nuclear Physics B 161~(2) (1979)
  493 -- 532.
\newblock \href
  {http://dx.doi.org/https://doi.org/10.1016/0550-3213(79)90225-6}
  {\path{doi:https://doi.org/10.1016/0550-3213(79)90225-6}}.

\bibitem{HALL1981397}
L.~J. Hall, M.~B. Wise, Flavor changing higgs boson couplings, Nuclear Physics
  B 187~(3) (1981) 397 -- 408.
\newblock \href
  {http://dx.doi.org/https://doi.org/10.1016/0550-3213(81)90469-7}
  {\path{doi:https://doi.org/10.1016/0550-3213(81)90469-7}}.

\bibitem{PhysRevD.41.3421}
V.~Barger, J.~L. Hewett, R.~J.~N. Phillips, New constraints on the charged
  higgs sector in two-higgs-doublet models, Phys. Rev. D 41 (1990) 3421--3441.
\newblock \href {http://dx.doi.org/10.1103/PhysRevD.41.3421}
  {\path{doi:10.1103/PhysRevD.41.3421}}.

\bibitem{Branco:2011iw}
G.~C. Branco, P.~M. Ferreira, L.~Lavoura, M.~N. Rebelo, M.~Sher, J.~P. Silva,
  {Theory and phenomenology of two-Higgs-doublet models}, Phys. Rept. 516
  (2012) 1--102.
\newblock \href {http://arxiv.org/abs/1106.0034} {\path{arXiv:1106.0034}},
  \href {http://dx.doi.org/10.1016/j.physrep.2012.02.002}
  {\path{doi:10.1016/j.physrep.2012.02.002}}.

\bibitem{Georgi:1985nv}
H.~Georgi, M.~Machacek, {DOUBLY CHARGED HIGGS BOSONS}, Nucl. Phys. B262 (1985)
  463--477.
\newblock \href {http://dx.doi.org/10.1016/0550-3213(85)90325-6}
  {\path{doi:10.1016/0550-3213(85)90325-6}}.

\bibitem{Chanowitz:1985ug}
M.~S. Chanowitz, M.~Golden, {Higgs Boson Triplets With M ($W$) = M ($Z$) $\cos
  \theta \omega$}, Phys. Lett. B 165 (1985) 105--108.
\newblock \href {http://dx.doi.org/10.1016/0370-2693(85)90700-2}
  {\path{doi:10.1016/0370-2693(85)90700-2}}.

\bibitem{Gunion:1989ci}
J.~Gunion, R.~Vega, J.~Wudka, {Higgs triplets in the standard model}, Phys.
  Rev. D 42 (1990) 1673--1691.
\newblock \href {http://dx.doi.org/10.1103/PhysRevD.42.1673}
  {\path{doi:10.1103/PhysRevD.42.1673}}.

\bibitem{Hartling:2014zca}
K.~Hartling, K.~Kumar, H.~E. Logan, {The decoupling limit in the
  Georgi-Machacek model}, Phys. Rev. D 90~(1) (2014) 015007.
\newblock \href {http://arxiv.org/abs/1404.2640} {\path{arXiv:1404.2640}},
  \href {http://dx.doi.org/10.1103/PhysRevD.90.015007}
  {\path{doi:10.1103/PhysRevD.90.015007}}.

\bibitem{Chiang:2015kka}
C.-W. Chiang, K.~Tsumura, {Properties and searches of the exotic neutral Higgs
  bosons in the Georgi-Machacek model}, JHEP 04 (2015) 113.
\newblock \href {http://arxiv.org/abs/1501.04257} {\path{arXiv:1501.04257}},
  \href {http://dx.doi.org/10.1007/JHEP04(2015)113}
  {\path{doi:10.1007/JHEP04(2015)113}}.

\bibitem{Chiang:2015rva}
C.-W. Chiang, S.~Kanemura, K.~Yagyu, {Phenomenology of the Georgi-Machacek
  model at future electron-positron colliders}, Phys. Rev. D 93~(5) (2016)
  055002.
\newblock \href {http://arxiv.org/abs/1510.06297} {\path{arXiv:1510.06297}},
  \href {http://dx.doi.org/10.1103/PhysRevD.93.055002}
  {\path{doi:10.1103/PhysRevD.93.055002}}.

\bibitem{Chiang:2018cgb}
C.-W. Chiang, G.~Cottin, O.~Eberhardt, {Global fits in the Georgi-Machacek
  model}, Phys. Rev. D 99~(1) (2019) 015001.
\newblock \href {http://arxiv.org/abs/1807.10660} {\path{arXiv:1807.10660}},
  \href {http://dx.doi.org/10.1103/PhysRevD.99.015001}
  {\path{doi:10.1103/PhysRevD.99.015001}}.

\bibitem{Das:2018vkv}
D.~Das, I.~Saha, {Cornering variants of the Georgi-Machacek model using Higgs
  precision data}, Phys. Rev. D 98~(9) (2018) 095010.
\newblock \href {http://arxiv.org/abs/1811.00979} {\path{arXiv:1811.00979}},
  \href {http://dx.doi.org/10.1103/PhysRevD.98.095010}
  {\path{doi:10.1103/PhysRevD.98.095010}}.

\bibitem{Chakrabortty:2013mha}
J.~Chakrabortty, P.~Konar, T.~Mondal, {Copositive Criteria and Boundedness of
  the Scalar Potential}, Phys. Rev. D 89~(9) (2014) 095008.
\newblock \href {http://arxiv.org/abs/1311.5666} {\path{arXiv:1311.5666}},
  \href {http://dx.doi.org/10.1103/PhysRevD.89.095008}
  {\path{doi:10.1103/PhysRevD.89.095008}}.

\bibitem{PING1993109}
L.~Ping, F.~Y. Yu, Criteria for copositive matrices of order four, Linear
  Algebra and its Applications 194 (1993) 109 -- 124.
\newblock \href
  {http://dx.doi.org/https://doi.org/10.1016/0024-3795(93)90116-6}
  {\path{doi:https://doi.org/10.1016/0024-3795(93)90116-6}}.

\bibitem{Gunion:1990dt}
J.~Gunion, R.~Vega, J.~Wudka, {Naturalness problems for rho = 1 and other large
  one loop effects for a standard model Higgs sector containing triplet
  fields}, Phys. Rev. D 43 (1991) 2322--2336.
\newblock \href {http://dx.doi.org/10.1103/PhysRevD.43.2322}
  {\path{doi:10.1103/PhysRevD.43.2322}}.

\bibitem{Blasi:2017xmc}
S.~Blasi, S.~De~Curtis, K.~Yagyu, {Effects of custodial symmetry breaking in
  the Georgi-Machacek model at high energies}, Phys. Rev. D96~(1) (2017)
  015001.
\newblock \href {http://arxiv.org/abs/1704.08512} {\path{arXiv:1704.08512}},
  \href {http://dx.doi.org/10.1103/PhysRevD.96.015001}
  {\path{doi:10.1103/PhysRevD.96.015001}}.

\bibitem{Keeshan:2018ypw}
B.~Keeshan, H.~E. Logan, T.~Pilkington, {Custodial symmetry violation in the
  Georgi-Machacek model}, Phys. Rev. D 102~(1) (2020) 015001.
\newblock \href {http://arxiv.org/abs/1807.11511} {\path{arXiv:1807.11511}},
  \href {http://dx.doi.org/10.1103/PhysRevD.102.015001}
  {\path{doi:10.1103/PhysRevD.102.015001}}.

\bibitem{Cynolter:2008ea}
G.~Cynolter, E.~Lendvai, {Electroweak Precision Constraints on Vector-like
  Fermions}, Eur. Phys. J. C58 (2008) 463--469.
\newblock \href {http://arxiv.org/abs/0804.4080} {\path{arXiv:0804.4080}},
  \href {http://dx.doi.org/10.1140/epjc/s10052-008-0771-7}
  {\path{doi:10.1140/epjc/s10052-008-0771-7}}.

\bibitem{He:2001tp}
H.-J. He, N.~Polonsky, S.-f. Su, {Extra families, Higgs spectrum and oblique
  corrections}, Phys. Rev. D64 (2001) 053004.
\newblock \href {http://arxiv.org/abs/hep-ph/0102144}
  {\path{arXiv:hep-ph/0102144}}, \href
  {http://dx.doi.org/10.1103/PhysRevD.64.053004}
  {\path{doi:10.1103/PhysRevD.64.053004}}.

\bibitem{Grimus:2007if}
W.~Grimus, L.~Lavoura, O.~M. Ogreid, P.~Osland, {A Precision constraint on
  multi-Higgs-doublet models}, J. Phys. G35 (2008) 075001.
\newblock \href {http://arxiv.org/abs/0711.4022} {\path{arXiv:0711.4022}},
  \href {http://dx.doi.org/10.1088/0954-3899/35/7/075001}
  {\path{doi:10.1088/0954-3899/35/7/075001}}.

\bibitem{Grimus:2008nb}
W.~Grimus, L.~Lavoura, O.~M. Ogreid, P.~Osland, {The Oblique parameters in
  multi-Higgs-doublet models}, Nucl. Phys. B801 (2008) 81--96.
\newblock \href {http://arxiv.org/abs/0802.4353} {\path{arXiv:0802.4353}},
  \href {http://dx.doi.org/10.1016/j.nuclphysb.2008.04.019}
  {\path{doi:10.1016/j.nuclphysb.2008.04.019}}.

\bibitem{Abbiendi:2013hk}
G.~Abbiendi, et~al., {Search for Charged Higgs bosons: Combined Results Using
  LEP Data}, Eur. Phys. J. C73 (2013) 2463.
\newblock \href {http://arxiv.org/abs/1301.6065} {\path{arXiv:1301.6065}},
  \href {http://dx.doi.org/10.1140/epjc/s10052-013-2463-1}
  {\path{doi:10.1140/epjc/s10052-013-2463-1}}.

\bibitem{Abdallah:2003xe}
J.~Abdallah, et~al., {Searches for supersymmetric particles in e+ e- collisions
  up to 208-GeV and interpretation of the results within the MSSM}, Eur. Phys.
  J. C31 (2003) 421--479.
\newblock \href {http://arxiv.org/abs/hep-ex/0311019}
  {\path{arXiv:hep-ex/0311019}}, \href
  {http://dx.doi.org/10.1140/epjc/s2003-01355-5}
  {\path{doi:10.1140/epjc/s2003-01355-5}}.

\bibitem{Achard:2001qw}
P.~Achard, et~al., {Search for heavy neutral and charged leptons in
  $e^{+}e^{-}$ annihilation at LEP}, Phys. Lett. B517 (2001) 75--85.
\newblock \href {http://arxiv.org/abs/hep-ex/0107015}
  {\path{arXiv:hep-ex/0107015}}, \href
  {http://dx.doi.org/10.1016/S0370-2693(01)01005-X}
  {\path{doi:10.1016/S0370-2693(01)01005-X}}.

\bibitem{Aaboud:2018xdt}
M.~Aaboud, et~al., {Measurements of Higgs boson properties in the diphoton
  decay channel with 36 fb$^{-1}$ of $pp$ collision data at $\sqrt{s} = 13$ TeV
  with the ATLAS detector}, Phys. Rev. D98 (2018) 052005.
\newblock \href {http://arxiv.org/abs/1802.04146} {\path{arXiv:1802.04146}},
  \href {http://dx.doi.org/10.1103/PhysRevD.98.052005}
  {\path{doi:10.1103/PhysRevD.98.052005}}.

\bibitem{Sirunyan:2018koj}
A.~M. Sirunyan, et~al., {Combined measurements of Higgs boson couplings in
  proton–proton collisions at $\sqrt{s}=13\,\text {Te}\text {V} $}, Eur.
  Phys. J. C79~(5) (2019) 421.
\newblock \href {http://arxiv.org/abs/1809.10733} {\path{arXiv:1809.10733}},
  \href {http://dx.doi.org/10.1140/epjc/s10052-019-6909-y}
  {\path{doi:10.1140/epjc/s10052-019-6909-y}}.

\bibitem{Arcadi:2018pfo}
G.~Arcadi, {2HDM portal for Singlet-Doublet Dark Matter}, Eur. Phys. J.
  C78~(10) (2018) 864.
\newblock \href {http://arxiv.org/abs/1804.04930} {\path{arXiv:1804.04930}},
  \href {http://dx.doi.org/10.1140/epjc/s10052-018-6327-6}
  {\path{doi:10.1140/epjc/s10052-018-6327-6}}.

\bibitem{Khachatryan:2014jba}
V.~Khachatryan, et~al., {Precise determination of the mass of the Higgs boson
  and tests of compatibility of its couplings with the standard model
  predictions using proton collisions at 7 and 8 $\,\text {TeV}$}, Eur. Phys.
  J. C75~(5) (2015) 212.
\newblock \href {http://arxiv.org/abs/1412.8662} {\path{arXiv:1412.8662}},
  \href {http://dx.doi.org/10.1140/epjc/s10052-015-3351-7}
  {\path{doi:10.1140/epjc/s10052-015-3351-7}}.

\bibitem{Aad:2015gba}
G.~Aad, et~al., {Measurements of the Higgs boson production and decay rates and
  coupling strengths using pp collision data at $\sqrt{s}=7$ and 8 TeV in the
  ATLAS experiment}, Eur. Phys. J. C76~(1) (2016) 6.
\newblock \href {http://arxiv.org/abs/1507.04548} {\path{arXiv:1507.04548}},
  \href {http://dx.doi.org/10.1140/epjc/s10052-015-3769-y}
  {\path{doi:10.1140/epjc/s10052-015-3769-y}}.

\bibitem{Bauer:2017fsw}
M.~Bauer, M.~Klassen, V.~Tenorth, {Universal properties of pseudoscalar
  mediators in dark matter extensions of 2HDMs}, JHEP 07 (2018) 107.
\newblock \href {http://arxiv.org/abs/1712.06597} {\path{arXiv:1712.06597}},
  \href {http://dx.doi.org/10.1007/JHEP07(2018)107}
  {\path{doi:10.1007/JHEP07(2018)107}}.

\bibitem{Amhis:2016xyh}
Y.~Amhis, et~al., {Averages of $b$-hadron, $c$-hadron, and $\tau$-lepton
  properties as of summer 2016}, Eur. Phys. J. C77~(12) (2017) 895.
\newblock \href {http://arxiv.org/abs/1612.07233} {\path{arXiv:1612.07233}},
  \href {http://dx.doi.org/10.1140/epjc/s10052-017-5058-4}
  {\path{doi:10.1140/epjc/s10052-017-5058-4}}.

\bibitem{Arbey:2017gmh}
A.~Arbey, F.~Mahmoudi, O.~Stal, T.~Stefaniak, {Status of the Charged Higgs
  Boson in Two Higgs Doublet Models}, Eur. Phys. J. C78~(3) (2018) 182.
\newblock \href {http://arxiv.org/abs/1706.07414} {\path{arXiv:1706.07414}},
  \href {http://dx.doi.org/10.1140/epjc/s10052-018-5651-1}
  {\path{doi:10.1140/epjc/s10052-018-5651-1}}.

\bibitem{Semenov:2008jy}
A.~Semenov, {LanHEP: A Package for the automatic generation of Feynman rules in
  field theory. Version 3.0}, Comput. Phys. Commun. 180 (2009) 431--454.
\newblock \href {http://arxiv.org/abs/0805.0555} {\path{arXiv:0805.0555}},
  \href {http://dx.doi.org/10.1016/j.cpc.2008.10.012}
  {\path{doi:10.1016/j.cpc.2008.10.012}}.

\bibitem{Belanger:2001fz}
G.~Belanger, F.~Boudjema, A.~Pukhov, A.~Semenov, {MicrOMEGAs: A Program for
  calculating the relic density in the MSSM}, Comput. Phys. Commun. 149 (2002)
  103--120.
\newblock \href {http://arxiv.org/abs/hep-ph/0112278}
  {\path{arXiv:hep-ph/0112278}}, \href
  {http://dx.doi.org/10.1016/S0010-4655(02)00596-9}
  {\path{doi:10.1016/S0010-4655(02)00596-9}}.

\bibitem{Aprile:2018dbl}
E.~Aprile, et~al., {Dark Matter Search Results from a One Tonne$\times$Year
  Exposure of XENON1T}\href {http://arxiv.org/abs/1805.12562}
  {\path{arXiv:1805.12562}}.

\bibitem{Aprile:2015uzo}
E.~Aprile, et~al., {Physics reach of the XENON1T dark matter experiment}, JCAP
  1604~(04) (2016) 027.
\newblock \href {http://arxiv.org/abs/1512.07501} {\path{arXiv:1512.07501}},
  \href {http://dx.doi.org/10.1088/1475-7516/2016/04/027}
  {\path{doi:10.1088/1475-7516/2016/04/027}}.

\bibitem{Tan:2016zwf}
A.~Tan, et~al., {Dark Matter Results from First 98.7 Days of Data from the
  PandaX-II Experiment}, Phys. Rev. Lett. 117~(12) (2016) 121303.
\newblock \href {http://arxiv.org/abs/1607.07400} {\path{arXiv:1607.07400}},
  \href {http://dx.doi.org/10.1103/PhysRevLett.117.121303}
  {\path{doi:10.1103/PhysRevLett.117.121303}}.

\bibitem{Cui:2017nnn}
X.~Cui, et~al., {Dark Matter Results From 54-Ton-Day Exposure of PandaX-II
  Experiment}\href {http://arxiv.org/abs/1708.06917} {\path{arXiv:1708.06917}}.

\bibitem{TheMEG:2016wtm}
A.~Baldini, et~al., {Search for the lepton flavour violating decay $\mu ^+
  \rightarrow \mathrm {e}^+ \gamma $ with the full dataset of the MEG
  experiment}, Eur. Phys. J. C 76~(8) (2016) 434.
\newblock \href {http://arxiv.org/abs/1605.05081} {\path{arXiv:1605.05081}},
  \href {http://dx.doi.org/10.1140/epjc/s10052-016-4271-x}
  {\path{doi:10.1140/epjc/s10052-016-4271-x}}.

\bibitem{Falkowski:2011xh}
A.~Falkowski, J.~T. Ruderman, T.~Volansky, {Asymmetric Dark Matter from
  Leptogenesis}, JHEP 05 (2011) 106.
\newblock \href {http://arxiv.org/abs/1101.4936} {\path{arXiv:1101.4936}},
  \href {http://dx.doi.org/10.1007/JHEP05(2011)106}
  {\path{doi:10.1007/JHEP05(2011)106}}.

\bibitem{DEramo:2017gpl}
F.~D'Eramo, N.~Fernandez, S.~Profumo, {When the Universe Expands Too Fast:
  Relentless Dark Matter}, JCAP 05 (2017) 012.
\newblock \href {http://arxiv.org/abs/1703.04793} {\path{arXiv:1703.04793}},
  \href {http://dx.doi.org/10.1088/1475-7516/2017/05/012}
  {\path{doi:10.1088/1475-7516/2017/05/012}}.

\bibitem{Dutta:2018zkg}
B.~Dutta, C.~S. Fong, E.~Jimenez, E.~Nardi, {A cosmological pathway to testable
  leptogenesis}, JCAP 10 (2018) 025.
\newblock \href {http://arxiv.org/abs/1804.07676} {\path{arXiv:1804.07676}},
  \href {http://dx.doi.org/10.1088/1475-7516/2018/10/025}
  {\path{doi:10.1088/1475-7516/2018/10/025}}.

\bibitem{Chen:2019etb}
S.-L. Chen, A.~Dutta~Banik, Z.-K. Liu, {Leptogenesis in fast expanding
  Universe}, JCAP 03 (2020) 009.
\newblock \href {http://arxiv.org/abs/1912.07185} {\path{arXiv:1912.07185}},
  \href {http://dx.doi.org/10.1088/1475-7516/2020/03/009}
  {\path{doi:10.1088/1475-7516/2020/03/009}}.

\bibitem{Mahanta:2019sfo}
D.~Mahanta, D.~Borah, {TeV Scale Leptogenesis with Dark Matter in Non-standard
  Cosmology}, JCAP 04~(04) (2020) 032.
\newblock \href {http://arxiv.org/abs/1912.09726} {\path{arXiv:1912.09726}},
  \href {http://dx.doi.org/10.1088/1475-7516/2020/04/032}
  {\path{doi:10.1088/1475-7516/2020/04/032}}.

\bibitem{Konar:2020vuu}
P.~Konar, A.~Mukherjee, A.~K. Saha, S.~Show, {A dark clue to seesaw and
  leptogenesis in singlet doublet scenario with (non)standard cosmology}\href
  {http://arxiv.org/abs/2007.15608} {\path{arXiv:2007.15608}}.

\end{thebibliography}

\end{document}